\RequirePackage{amsmath}
\documentclass[runningheads]{llncs}
\usepackage{graphicx}
\usepackage{tikz}
\usetikzlibrary{calc}
\tikzset{
	axes/.style={thick, ->},
	axes2/.style={thick},
	fd/.style={color=red},
	gfd/.style={color=blue},
	g2fd/.style={color=green},
	g3fd/.style={color=cyan}
}

\usepackage{amssymb}
\usepackage{algorithm, algpseudocode}
\usepackage{enumitem}
\usepackage{hyperref}
\usepackage[capitalise]{cleveref}
\usepackage{fullpage}
\DeclareMathOperator{\fix}{fix}
\newcommand{\Lex}{\text{Lex}}

\newcommand{\lexx}{\succ}
\newcommand{\N}{\mathbb{N}}

\newcommand{\R}{\mathbb{R}}

\newcommand{\talque}{\: : \:}

\newcommand\xqed{%
  \leavevmode\unskip\penalty9999 \hbox{}\nobreak\hfill
  \quad\hbox{$\triangle$}}

\DeclareMathOperator{\squad}{\hspace{0.5em}} %small \quad
\DeclareMathOperator*{\argmax}{argmax}
\DeclareMathOperator{\id}{id}
\DeclareMathOperator{\im}{Im}
\DeclareMathOperator{\inte}{int}
\DeclareMathOperator{\Orb}{Orb}
\DeclareMathOperator{\Ort}{O}
\DeclareMathOperator{\SSP}{\text{SS}}

\DeclareMathOperator{\interior}{int}
\DeclareMathOperator{\relint}{relint}

\begin{document}
\title{On the Geometry of Symmetry Breaking Inequalities}%\thanks{A preliminary version of this manuscript appeared in the proceedings of IPCO 2021~\cite{verschae+etal21}.}}

\author{
    José Verschae\inst{1}
    \and Matías Villagra\inst{2}% \orcidID{1111-2222-3333-4444} 
    \and Léonard von Niederhäusern\inst{3}
    %José Verschae\inst{1}\orcidID{0000-0002-2049-6467} 
    %\and Matías Villagra\inst{2}% \orcidID{1111-2222-3333-4444} 
    %\and Léonard von Niederhäusern\inst{3}\orcidID{0000-0001-6113-8239}
} % Alphabetical order ? 
%
%\authorrunning{J. Verschae et al.}
\institute{Pontificia Universidad Católica, Institute for Mathematical and Computational Engineering \\ Faculty of Mathematics and School of Engineering, Chile \email{jverschae@uc.cl}\and
Pontificia Universidad Católica de Chile, Faculty of Mathematics, Chile and  \\ Columbia University, IEOR, USA \email{mjv2153@columbia.edu} \and
Universidad de O'Higgins, Institute for Engineering Sciences and \\ Universidad de Chile, Center for Mathematical Modelling (AFB170001),  Chile\\
\email{leonard.vonniederhausern@uoh.cl}}

\maketitle   

\begin{abstract}
Breaking symmetries is a popular way of speeding up the branch-and-bound method for symmetric integer programs. We study fundamental domains, which are minimal and closed symmetry breaking polyhedra. Our long-term goal is to understand the relationship between the complexity of such polyhedra and their symmetry breaking capability.

Borrowing ideas from geometric group theory, we provide structural properties that relate the action of the group with the geometry of the facets of fundamental domains. Inspired by these insights, we provide a new generalized construction for fundamental domains, which we call generalized Dirichlet domain (GDD). Our construction is recursive and exploits the coset decomposition of the subgroups that fix given vectors in $\mathbb{R}^n$. We use this construction to analyze a recently introduced set of symmetry breaking inequalities by Salvagnin~\cite{salvagnin18} and Liberti and Ostrowski~\cite{liberti+ostrowski14}, called Schreier-Sims inequalities. In particular, this shows that every permutation group admits a fundamental domain with less than $n$ facets. We also show that this bound is tight. 

Finally, we prove that the Schreier-Sims inequalities can contain an exponential number of isomorphic binary vectors for a given permutation group~$G$, which provides evidence of the lack of symmetry breaking effectiveness of this fundamental domain. Conversely, a suitably constructed GDD for this $G$ has linearly many inequalities and contains unique representatives for isomorphic binary vectors.

\end{abstract}

\section{Introduction}

Symmetries are mappings from one object into itself that preserve its structure. Their study has proven fruitful across a myriad of fields, including integer programming, where symmetries are commonly present. For instance, almost 30\% of mixed-integer linear programs (MILP) in the model library used by the solver CPLEX are considerably affected by symmetry~\cite{achterberg+etal13}. Moreover, symmetry exploitation techniques are of importance in various situations. In particular, they help to avoid traversing symmetric branches of the tree considered by a branch-and-bound algorithm.

Roughly speaking, the symmetry group $G$ of an optimization problem is the set of functions in $\mathbb{R}^n$ that leave the feasible region and the objective function invariant (see Section 2 for a precise definition). The symmetry group $G$, or any of its subgroups, partitions $\mathbb{R}^n$ into $G$-\emph{orbits}, which are sets of isomorphic solutions. A natural technique for handling symmetries is to add a static set of symmetry breaking inequalities. That is, we add extra inequalities that remove isomorphic solutions while leaving at least one representative per $G$-orbit. This well established approach has been studied extensively, both in general settings and different applications; see e.g.~\cite{faenza2009extended,friedman07,ghoniem+sherali11,hojny+pfetsch18,kaibel+pfetsch06,liberti12a,liberti+ostrowski14,ostrowski2010symmetry,salvagnin18,sherali+smith01}. In most of these works, the symmetry breaking inequalities select the lexicographically maximal vector in each $G$-orbit of binary vectors. However, this constitutes a major drawback when dealing with general permutation groups: selecting the lexicographically maximal vector in a $G$-orbit is an NP-hard problem~\cite{babai+luks83}. Hence, the separation problem of the corresponding symmetry breaking inequalities is also NP-hard. On the other hand, there is nothing preventing us to select orbit representatives with a different criterion.

In this article, we are interested in understanding \emph{fundamental domains} of a given finite group $G$, which are minimal, closed and convex symmetry breaking sets for $G$. %  are closed and convex sets of $\R^n$ that
Ideally, a closed symmetry breaking set $F$ contains a unique representative per $G$-orbit. However, such a set does not necessarily exist for every group. Instead, a fundamental domain $F$ only contains a unique representative for $G$-orbits that intersect $F$ in its interior, while it can contain one or more representatives of a $G$-orbit intersecting its boundary. Despite this, $F$ is a minimal closed symmetry breaking set, as any proper closed subset of $F$ leaves some $G$-orbit unrepresented. On the other hand, a given symmetry group can admit inherently different fundamental domains. While all fundamental domains for finite orthogonal groups, including permutation groups (the main focus when considering mixed integer linear programs), are polyhedral cones, their polyhedral structure and complexity might differ greatly.

Our long term and ambitious goal is to understand the tension (and potential trade-offs) between the symmetry breaking effectiveness and the complexity of fundamental domains. The complexity can be measured in several ways: from the sizes of the coefficients in its matrix description, the number of facets, or even its extension complexity. On the other hand, the symmetry breaking effectiveness is related to the number of representatives that each orbit contains. Hence, the boundary of a fundamental domain, which can contain overrepresented $G$-orbits, becomes problematic, in particular if our points of interest (e.g., binary points in a binary integer program) can lie within it. %This is surprisingly common in constructions of fundamental domains, as we will see.

More precisely, we contribute to the following essential questions: (i) Which groups admit fundamental domains in $\mathbb{R}^n$ with $\text{poly}(n)$ facets? (ii) What is the structure of these facets? (iii) Which algorithmic methods can we use to construct different fundamental domains? (iv) Which fundamental domains contain unique representatives for every orbit? 

\paragraph{Related Work.} The concept of fundamental domain traces back to the 19th century, as it corresponds to fundamental parallelepipeds for the symmetry group of a lattice. Fundamental domains are studied in several areas, for example crystallography, the theory of quadratic forms, and elliptic functions, among many others. 
In particular Dirichlet~\cite{dirichlet1850} gives a construction which implies the existence of a fundamental domain in a general context, including all groups of isometries in $\mathbb{R}^n$, later known as \emph{Dirichlet domain}. For a historical overview see Ratcliffe~\cite{ratcliffe19} and the references therein.
%This concept can be generalized even for any \emph{Fuchsian group}. 
%, including historical references.

Several techniques have been studied to handle symmetries in integer programming. Kaibel and Pfetsch \cite{kaibel+pfetsch06} introduce the concept of \emph{orbitopes} as the convex-hull of $0-1$ matrices that are lexicographically maximal under column permutations, and give a complete description of the facets for the cyclic group and the symmetric group. Friedman~\cite{friedman07} considers general permutation groups. Based on the Dirichlet Domain, he introduces the idea of a universal ordering vector, which yields a fundamental domain with unique representatives of binary points. On the other hand, this fundamental domain has an exponential number of facets, its defining inequalities can contain exponentially large coefficients in $n$, and the separation problem is NP-hard for general permutation groups~\cite{babai+luks83,margot10}. Liberti~\cite{liberti12a} and later Dias and Liberti~\cite{dias+liberti19} also consider general permutation groups $G$ and derive a class of symmetry breaking constraints by studying the orbits of $G$ acting on $[n]=\{1,\ldots,n\}$. Liberti and Ostrowski~\cite{liberti+ostrowski14}, and independently Salvagnin~\cite{salvagnin18}, extend this construction and introduce a set of symmetry breaking inequalities based on a chain of pointwise coordinate stabilizers. We will refer to this set as the Schreier-Sims inequalities, as they are strongly related to the \emph{Schreier-Sims table} from computational group theory~\cite{seress03}. %In Constraint Programming a similar construction has been used by Crawford et al.~\cite{crawford+etal96} and Puget~\cite{puget03}.
Hojny and Pfetsch~\cite{hojny+pfetsch18} study \emph{symretopes}, defined as the convex hulls of lexicographically maximal vectors in binary orbits. %That is, the convex hull of the fundamental domain defined by Friedman~\cite{friedman07}. 
They obtain a linear time algorithm for separating the convex hull of polytopes derived by a single lexicographic order enforcing inequality and show how to exploit this construction computationally.

For integer programming techniques, dynamic methods have been used to deal with symmetries within the Branch-and-Bound tree. Some methods are \emph{Orbital Fixing}~\cite{margot03}, \emph{Isomorphism Pruning}~\cite{margot02} and \emph{Orbital Branching}~\cite{ostrowski11}. A more geometric approach for solving symmetric integer programs relies on the theory of \emph{core points}~\cite{bodi+etal13,herr+etal13}. For more details on these techniques and related topics see Margot~\cite{margot10}, Pfetsch and Rehn~\cite{pfetsch_computational_2019}, and Schürmann~\cite{schurmann13}. %provides an interesting overview of symmetry exploitation techniques for areas beyond integer programming.

\paragraph{Our Contribution.} In this article we focus on finite orthogonal groups in $\R^{n}$, that is, groups of linear isometries. We start by presenting basic structural results of the theory of fundamental domains for a given orthogonal group $G$. A basic observation is that each facet is related to a group element $g$. We also show the following new property of the facets: for an interesting class of fundamental domains, which we call \emph{subgroup consistent}, the vector defining a facet must be orthogonal to the fixed subspace of $g$. This implies that each inequality is of the form $\alpha^t x\ge \alpha^t (gx)$ for some vector $\alpha\in \mathbb{R}^n$ and some element $g\in G$. In other words, the inequalities of any subgroup consistent fundamental domain have the same structure as inequalities of Dirichlet domains. 

Inspired by these new insights, we state our main contribution: a generalized construction of fundamental domains for any finite orthogonal group, including permutation groups. Our method is based on choosing a vector $\alpha$ and finding the coset decomposition using the stabilizer subgroup $G_{\alpha}=\{g\in G\talque g\alpha = \alpha\}$. %We add inequalities to our fundamental domain, one for each member in the coset decomposition. Therefore, for a well chosen $\alpha$ the number of cosets can be bounded, yielding a polynomial number of inequalities. Then, we proceed recursively on the subgroup $G_{\alpha}$. 
Next, we add inequalities to our symmetry breaking set, one for each member in the coset decomposition. For a  well-chosen $\alpha$, the number of cosets can be bounded by a polynomial, yielding a polynomial number of inequalities. By proceeding recursively on the subgroup $G_{\alpha}$, we generate a fundamental domain after at most $n$ iterations.
We say that a fundamental domain obtained via this method is a \emph{generalized Dirichlet domain} (GDD), as it generalizes the classical construction by Dirichlet~\cite{dirichlet1850}. To the best of our knowledge, this construction generalizes all convex fundamental domains found in the literature. For the special case of permutation groups, our algorithm can be implemented in polynomial time if the vector $\alpha$ is well chosen.

A natural way of breaking symmetries is to choose the lexicographically maximal element for \emph{every} $G$-orbit in $\mathbb{R}^n$ (not only binary vectors, as in the construction by Friedman~\cite{friedman07}). However, it is not hard to see that the obtained set is not necessarily closed. On the other hand, the set is convex. We show that the closure of this set coincides with the Schreier-Sims inequalities studied by Salvagnin~\cite{salvagnin18} and Liberti and Ostrowski~\cite{liberti+ostrowski14}. Moreover, we show that this set is a GDD, which implies that it is a fundamental domain. Finally, we give a stronger bound on the number of facets for this fundamental domain, implying that all permutation groups admit a fundamental domain with at most $n-1$ inequalities. We also notice that any fundamental domain for $S_{n}$, the full symmetric group of degree $n$, has $n-1$ facets, which shows that our bound is best possible.

Salvagnin~\cite{salvagnin18} recognizes that the symmetry breaking efficiency of the Schreier-Sims inequalities might be limited: the orbit of a binary vector can be overrepresented in the set. We give a specific example of a permutation group in which an orbit of binary vectors can have up to $2^{\Omega(n)}$ many representatives. Using the flexibility given by our GDD construction, we exhibit a fundamental domain for the same group with a unique representative for each binary orbit, while having $O(n)$ facets. This illustrates that exploiting the structure of the given group can yield a relevant improvement in the way symmetries are broken. Moreover, we show that the only groups that admit a fundamental domain with a unique representative for \emph{every} orbit are reflection groups. Finally, we propose a new way of measuring the effectiveness of fundamental domains, which we hope will pave the road for future work in deriving fundamental domains that exploit the structure of the groups involved. %Due to space reasons we refer several proofs and details to the full version of this paper~\cite{verschae2020geometry}. 

\section{Notation and Preliminaries}
% Preliminaries: 1.5 pages

Throughout the whole paper, $G$ denotes a group, and $H\leq G$ means that $H$ is a subgroup of $G$. The element $\id\in G$ denotes the identity. For a subset $S$ of $G$, $\langle S \rangle$ is the smallest group containing $S$. The set $\Ort_n(\R)$ corresponds to the orthogonal group in $\mathbb{R}^n$, that is, the group of all $n\times n$ orthogonal matrices (equivalently, linear isometries). Hence, it holds that if $g\in \Ort_n(\R)$ then the inverse $g^{-1}$ equals the transpose $g^t$. All groups considered in what follows are finite subgroups of $\Ort_n(\R)$. 
Also, $G_{(S)}$ denotes the pointwise stabilizer of the set $S\subseteq\R^n$, and the set $\text{fix}(g)$ denotes the invariant subspace of $g\in G$, that is, 
\[
    G_{(S)} := \left\{ g\in G \talque x = gx \squad \forall x\in S \right\} \text{ and } \text{fix}(g):=\left\{ x\in \R^n \talque gx = x \right\}.
\]
If $S :=\{x\}$, we write $G_{x} : = G_{(S)}$.
%Also, $G_{(S)}$ denotes the pointwise stabilizer of the set $S$, that is $G_{(S)} := \left\{ g\in G \talque x = gx \squad \forall x\in S \right\}.$ If $S :=\{x\}$, we write $G_{x} : = G_{(S)}$. The set $\text{fix}(g)$ denotes the invariant subspace of $g\in G$, i.e. $\text{fix}(g):=\left\{ x\in \R^n \talque gx = x \right\}.$
For $H\leq G$, a \emph{transversal} for $H$ in $G$ is a set of representatives from the left cosets of $H$ in $G$, the set of left cosets being $\{ gH\talque g\in G \}$. Given a set of elements $S\subseteq G$, we denote by $S^{-1} := \{g^{-1} \talque g\in S\}$.
For $x\in\R^n$, the $G$-orbit of $x$ is the set $\Orb_G(x) := \{ gx \talque g\in G\}.$ We denote by $[n]$ the set $\{1,\dots,n\}$ for all $n\in\N$ and $S_n$ denotes the symmetric group, that is the group of all permutations over $[n]$. For $G\leq S_n$, each element $g\in G$ acts on $\R^n$ by the mapping $x \mapsto gx := \left( x_{g^{-1}(i)} \right)_{i=1}^n$. Equivalently, we consider $G\leq S_n$ as a group of isometries where each $g\in G$ is interpreted as the corresponding permutation matrix.

For an exhaustive introduction to group theory, see for instance Rotman~\cite{rotman95}. For an exposition on computational aspects of permutation groups, see Seress~\cite{seress03}.

% Defined later and needed: signification of \langle S \rangle (group generated by S ?), notation G_([k]), Ort_n(\R)

For a set $S$ we denote by $S^c$ its complement. For $S\subseteq \mathbb{R}^n$, we write $\inte(S)$ for its interior, $\overline{S}$ for its closure, and $\partial S$ for its boundary. 

An optimization problem $\min\{f(x) \talque x\in X\}$ is \emph{$G$-invariant} if for all feasible $x$ and $g\in G$, 
\begin{enumerate}
    \item $f(x)=f(gx)$, and
    \item $gx$ is feasible.
\end{enumerate}

Given a $G$-invariant optimization problem, we can use the group $G$ to restrict the search of solutions to a subset of $\mathbb{R}^n$, namely a \emph{fundamental domain}.

\begin{definition}
A subset $F$ of $\R^n$ is a \emph{fundamental domain} for $G\leq \Ort_n(\mathbb{R})$ if 
\begin{enumerate}
    \item the set $F$ is closed and convex\footnote{Notice that in part of the literature, e.g.~\cite{ratcliffe19}, convexity is not part of the definition.},
    \item the members of $\left\{\inte(gF)\talque g\in G \right\}$ are pairwise disjoint,
    \item $\R^n = \bigcup_{g\in G} gF$.
\end{enumerate}
%If $F$ is not connected, it is said to be a fundamental \emph{region} for $G$.
\end{definition}
 Notice that for any $x \in \R^{n}$, its $G$-orbit, $\Orb_G(x)$, satisfies that $|\Orb_G(x) \cap F|\ge 1$. Also, $|\Orb_G(x) \cap F|= 1$ if $x\in \inte{(F)}$. It is not hard to see that all fundamental domains for a finite subgroup of $\Ort_n(\mathbb{R})$ are full-dimensional sets. Moreover, if $F'\subsetneq F$, then there is some $\Orb_G(x)$ such that $\Orb_G(x)\cap F' = \emptyset$, and hence some orbit is not represented in $F$. %We also remark that $F$ must be full-dimensional since $\R^n = \bigcup_{g\in G} gF$ and $G$ is finite.
 
 %contains (at least) a representative of the orbit $\Orb_G(x) = \left\{gx \talque g\in G\right\}$ for each $x$ in the domain.

\begin{definition}
A subset $R$ of $\R^n$ is a \emph{fundamental set} for a group $G\le \Ort_n(\R)$ if it contains exactly one representative of each $G$-orbit in $\R^n$.
\end{definition}

 \section{The Geometric Structure of Fundamental Domains}

% Geometric structure: 2 pages
In this section we review some basic geometric properties of fundamental domains and derive new properties. Propositions~\ref{subsection_fd_existence} and~\ref{prop:sidepairing} are well known; their proof can be found in~\cite[Ch. 6]{ratcliffe19}. Proposition~\ref{proposition:fd_generators} extends a similar result for the  particular case of \emph{exact} fundamental domains~\cite[Ch. 6]{ratcliffe19}. Theorem~\ref{thm:FD_facet_subgroup_consistent} and Corollary~\ref{cor:pairs} are our main contributions of this section. To provide a self-contained presentation of the topic, we provide alternative proofs of some of the previously known results.

%The first proposition of this section states that for every finite orthogonal group there exists a convex fundamental domain in $\R^{n}$. Even more, this fundamental domain is a polyhedral cone. Then we study a useful characterization of its facets and conclude the subsection with a result that beautifully connects the geometry of this polyhedron and its group structure; the facets of this fundamental polyhedron constitute a set of generators for the group. \ref{subsection_fd_existence} and \ref{proposition:fd_generators} were already known but we provide alternative proofs. We start by presenting three useful lemmas. 

The following proposition, together with the existence of a vector $\alpha$ whose stabilizer is trivial~\cite[Thm. 6.6.10.]{ratcliffe19}, guarantees the existence of a fundamental domain for any $G\leq \Ort_n(\mathbb{R})$. We will refer to the construction $F_{\alpha}$ in the proposition as a \emph{Dirichlet domain}.

\begin{proposition}
\label{subsection_fd_existence}
Let $G \leq \Ort_{n}(\R)$ be finite and non-trivial, and let $\alpha\in\R^{n}$ whose stabilizer $G_{\alpha}$ equals $\{\id\}$. Then the following set is a fundamental domain for~$G$,
\begin{equation*}
    F_{\alpha} = \{ x \in \R^{n} \talque \alpha^{t} x \geq \alpha^{t} g x, \squad \forall g \in G \}.
\end{equation*}
\end{proposition}
\begin{proof}
Let $\alpha \in \R^{n}$ be a point such that $G_{\alpha} = \langle \id \rangle$. Consider the linear functional $y \mapsto \alpha^{t}y$, for all $y \in \R^{n}$. Note that $z \in F_{\alpha}$ if and only if $z$ maximizes this linear functional over its finite $G$-orbit, i.e. $z \in \argmax \{ \alpha^{t}y \talque y \in \Orb_{G}(z) \}$. %We show that $F_{\alpha}$ is a fundamental domain for $G$.

First note that $F_{\alpha}$ is closed and convex by construction. Now, let $x \in \R^{n}$, and suppose that $z \in \argmax \{ \alpha^{t}y \talque y \in \Orb_{G}(x) \}$, and hence $z\in F_{\alpha}$. Then there exists a $g \in G$ such that $z = gx$, i.e. $x = g^{-1}z$. Hence $x \in g^{-1} F_{\alpha}$. Now, note that
\begin{align*}
    \inte (F_{\alpha}) &= \{ x \in \R^{n} \talque \alpha^{t} x > \alpha^{t} g x, \squad \forall g \in G \setminus \{\id \} \} \\
    &= \{ x \in \R^{n} \talque G_{x} = \{\id\}  \text{ and for all } y \in \Orb_{G}(x) \setminus \{ x \}, \squad \alpha^{t} x > \alpha^{t} y \}.
\end{align*}
Since every $g \in G$ is a linear homeomorphism, we have that $\inte (g F_{\alpha}) = g \inte (F_{\alpha})$. Thus, if $x \in \inte ( F_{\alpha})$ and $x \in g \inte (F_{\alpha})$, for some $g \in G$ non-trivial, there exists $y \in \inte (F_{\alpha})$, such that $x = g y$. But this is a contradiction since the maximum is unique and $G_{x}$ is trivial. In consequence, $F_{\alpha}$ is a fundamental domain for $G$.
% 
% Finally if we write $F_{\alpha} =  \{ x \in \R^{n} \talque \alpha^{t}(\id - g) x \geq 0, \squad \forall g \in G \}$, we notice that $F_{\alpha}$ is defined by a set of homogeneous linear inequalities, and hence is a (convex) polyhedral cone. 
\qed
\end{proof}

A specific kind of Dirichlet domains are \emph{$k$-fundamental domains}. For any integer $k\geq 2$, we define $\overline{k} := \left(k^{n-1}, k^{n-2}, \dots, 1\right)$ as the \emph{$k$-universal ordering vector}. 
The set $F_{\overline{k}}$ is the $k$-fundamental domain for the symmetry group $G$. Friedman~\cite{friedman07} observes that $F_{\overline{2}}$ contains a unique representative per $G$-orbit of binary points in $\R^n$. This fact easily generalizes for points $x\in\{0,\dots,k-1\}^n$ with the $k$-ordering vector (see \cite{margot10}).

Given a fundamental domain $F$ and $g \in G \setminus \{ \id \}$, let $H_{g}$ be any closed half-space that separates~$F$ and~$gF$, that is, $F \subseteq H_{g}$ and $gF \subseteq \overline{H_{g}^{c}}$. The existence of this half-space follows from % the fact that $\inte (F)$ and $\inte(gF)$ are disjoint convex sets and 
the convex separation theorem. We say that a collection $\{H_g \}_{g\in G}$ \emph{represents} $F$ if for every $g\in G$, the set $H_g$ is a closed half-space that separates $F$ and $gF$. Notice that representations are non unique.

Let us denote by $H_{g}^=:=\partial(H_g)$ the hyperplane defining $H_g$. Notice that $H_g^=$ contains $0$ as $0\in F\cap gF$, since $g$ is a linear isometry. %Since $F$ is a polyhedral cone, for every facet $S=F\cap gF$ there exists a vector $\gamma_{g} \in \R^{n}$ defining it.
We let $\gamma_g\neq 0$ be some defining vector for $H_g$, i.e., $H_g = \{x \in \R^{n} \talque \gamma_g^tx\ge0\}$, and thus $H_g^=  = \{x \in \R^{n} \talque \gamma_g^tx=0\}$. 

% An interesting connection between the geometry of fundamental domains and their underlying algebraic structure. This was already known in the literature and Ratcliffe~\cite{ratcliffe19} presents a similar proof for discrete groups of isometries of Euclidean and non-Euclidean metric spaces.

\begin{proposition}\label{proposition:fd_generators}
Let $G \leq \Ort_{n}(\R)$ be finite, let $F$ be a fundamental domain and  $\{H_g \}_{g\in G}$ a collection that represents $F$. Then $F=\bigcap_{g \in G} H_g$. In particular, $F$ is a polyhedral cone. Moreover, if $A$ is the set of all $g\in G$ such that ${\text{dim}(F\cap gF)=n-1}$, then $A$ generates $G$.
\end{proposition}
\begin{proof}
Let $H:=\bigcap_{g\in G} H_{g}$. Let us first show that $F= H.$
Clearly $F \subseteq H$ as $F\subseteq H_g$ for every $g\in G$. For the other inclusion, suppose by contradiction that $H \setminus F \neq \emptyset$. Given that $H$ is convex, $\inte(H)\neq \emptyset$ (as $F\subseteq H$ is full-dimensional), and $F^c$ is open, we have that $\inte (H \setminus F) \neq \emptyset$. %Indeed, if this were not the case, any point in $H \setminus F$ would be a boundary point, and so there would exist a sequence $\{ x_n \}_{n \in \N} \subseteq F$ converging to this point, which is absurd. 
Hence, let $x \in \inte (H \setminus F)$. As $F$ is a fundamental domain for $G$, there exists a $g \in G$ such that $gx \in F$. This implies that $x \in g^{-1} F \subseteq \overline{H_{g^{-1}}^{c}}$. By definition, we also have that $x \in H \subseteq H_{g^{-1}}$, thus $x \in H_{g^{-1}}^=$. But this contradicts the fact that $x$ belongs to the interior of $H_{g^{-1}}$, because $x$ belongs to the interior of $H \setminus F$. 

Let us now show that $A$ generates $G$. For a fixed $g\in G$, let us prove that $g \in \langle A \rangle$. First, take $x \in \inte(F)$ and $y\in \inte(gF)$. Now, take $\epsilon>0$ so that $B_{\epsilon}(x) \subseteq \inte(F)$, and choose $x_0 \in B_{\epsilon}(x)$ uniformly at random. For $a,b\in\mathbb{R}^n$, let $[a,b]$ denote the interval $\{\lambda a + (1-\lambda)b\talque \lambda\in [0,1]\}$. The interval $[x_0,y]$ is partitioned in several segments by the tessellation $\{hF\}_{h\in G}$. More precisely, notice that
\begin{equation}
\label{eq:intervalTessellation}
  [x_0,y] = \bigcup_{h\in G} ([x_0,y]\cap hF).
\end{equation}
As $hF$ is closed and convex, the set $[x_0,y]\cap hF$ is a  (possible empty) closed interval. Let $\lambda_0:=0$ and $g_0:=\id$. Define $\lambda_1$ as the maximum value such that $x_1:=\lambda_1x_0+(1-\lambda_1)y\in F$. Hence, $[x_0,x_1]\subseteq F$. By~\eqref{eq:intervalTessellation}, there must exist an element $g_1\neq \id$ such that $x_1\in g_1F$. More generally, given $\lambda_i\in(0,1)$, $g_i\neq g$, and $x_i\in g_iF$, let $\lambda_{i+1}$ be the maximum number such that $x_{i+1}:= \lambda_{i+1}x_0+(1-\lambda_{i+1})y\in g_iF$. As before, there must exists $g_{i+1}\in G\setminus\{g_0,g_1,\ldots,g_i\}$ such that $x_{i+1}\in g_{i+1}F$. The construction finishes as $G$ is finite, when we reach that $x_m\in gF$ for some $m$. Defining $x_{m+1}=y$ and $g_m=g$ we obtain that
\[
[x_0,y] = \bigcup_{i=1}^{m+1} [x_{i-1},x_i],
\]
where $[x_{i-1},x_i]\subseteq g_{i-1}F$ for all $i\in\{1,\ldots,m+1\}$. 

By construction, $x_i \in g_{i-1}F\cap g_{i}F$ for all $i\in\{1,\ldots,m\}$. %Equivalently, it holds that $g_{i-1}^{-1}x_i\in F \cap g_{i-1}^{-1}g_i F$. 
Moreover, we have the following claim.

\noindent\textbf{Claim 1:} It holds almost surely (a.s.) that for all $i\in\{1,\ldots,m\}$ the set $g_{i-1}F\cap g_{i}F$ has dimension $n-1$.

Let us show the claim. If the claim is not true, there must exists $i$ such that ${ \mathbb{P}( \text{dim} (g_{i-1}F\cap g_{i}F) = n-1 ) }$ with non-zero probability. For $h,h'\in G$, let $E(i,h,h')$ be the event that $g_{i-1}=h$ and $g_{i}=h'$. We will show that if $\text{dim}(hF\cap h'F)\le n-2$ then the probability of $E(i,h,h')$ is 0. This suffices to show the claim as $G$ is finite.
 
In the event $E(i,h,h')$, $x_i$ belongs to $R=hF\cap h'F$, where $\text{dim}(R)\leq n-2$. Notice now that
\begin{align*}
B'(R)&:=\{z\in B_{\epsilon}(x)\talque [z,y]\cap R\neq \emptyset \}\\ &= \left\lbrace z\in B_{\epsilon}(x)\talque z= \frac{1}{t}r + \frac{t-1}{t}y, r\in R, t\in (0,1)\right\rbrace \subseteq \text{affine}(R\cup \{y\}),
\end{align*} 
where $\text{affine}(S)$ denotes the affine span of $S$. Hence, $\dim(B'(R))\le \dim(\text{affine}(R\cup \{y\}))\le n-1$. This implies that the probability of $E(i,h,h')$ is 0, and hence the claim follows.%We conclude that, for a given $R$, the event $x_0\in B'(R)$ has 0 probability. Moreover, the probability of the event  $x_0\in \bigcup_{R}B'(R)$ is 0, where the union is taken over all facets of $hF$ of dimension at most $n-2$, for some $h\in G$; a finite set. We conclude that the event $A_i$ has 0 probability and hence also $\cup_{i=1}^m A_i$. This implies Claim 1.

By Claim 1 we know that for all $i\in\{1,\ldots,m\}$ the set $F\cap g_{i-1}^{-1}g_{i}F$ has dimension $n-1$ almost surely. Hence, we conclude the following.

\noindent\textbf{Claim 2:} For all $i\in\{1,\ldots,m\}$ it holds that $g_{i-1}^{-1}g_i\in  A$ almost surely.

We now conclude the theorem from this claim. Let us pick a sequence $g_0,\ldots,g_m$ that satisfies Claim 2, which exists as the claimed event has non-zero probability. Clearly, $g_0 = \id \in \langle A \rangle$. Moreover, $g_i = g_{i-1}h$ for some $h\in A $. Hence, if $g_{i-1}\in \langle A \rangle$, we have that $g_i\in \langle A \rangle$ for all $i$. Inductively, we obtain that $g=g_m\in \langle A \rangle$. \qed

\end{proof}

We now introduce a new type of fundamental domain and characterize the structure of its facets.

\begin{definition}
 A fundamental domain $F$ is said to be \emph{subgroup consistent for the collection} $\{H_g\}_{g\in G}$ representing $F$ if for every subgroup $G'\leq G$ the set $F'=\bigcap_{g \in G'} H_g$ is a fundamental domain for $G'$. We say that $F$ is \emph{subgroup consistent} if $F$ is subgroup consistent for some collection $\{H_g\}_{g\in G}$.
\end{definition}

It is not hard to see that Dirichlet domains are subgroup consistent. Moreover, subgroup consistent fundamental domains are amenable to be constructed iteratively, either by starting the construction of a fundamental domain for a subgroup and extending it to larger subgroups (bottom-up), or adding inequalities for $G$ and recurse to smaller subgroups (top-down, as our technique in Section~\ref{sec:GDD}).

%such that for every $x \in F \cap gF$,
%\begin{equation*} 
%    \gamma_{g}^{t}x = 0,  \quad \text{and} \quad F \subseteq H_{\gamma_{g}}^{\geq}.
%\end{equation*}
%In other words, $\gamma_{g}$ is a normal vector of the supporting hyperplane $H_{\gamma_{g}}$ at any $x \in F \cap gF$. 

With the help of the following lemmas, we show a close relationship between supporting hyperplanes of a subgroup consistent fundamental domain $F$: all facet-defining inequalities of $F$ are of the form $\alpha^tx \ge \alpha^t gx$ for some $\alpha$ and $g\in G$. In this case we say that the inequality is of \emph{Dirichlet type}. 

%To the best of our knowledge this property of facets is new.

\begin{lemma}\label{lemma:fix_interF}
Let $g \in G$. Then $(\fix(g) \cap F) \setminus H_{g}^{=} = \emptyset$.
\end{lemma}
\begin{proof}
Let $x \in \fix(g)$. If $x \in F \setminus H_{g}^{=}$, then $\gamma_{g}^{t} x > 0$. Moreover, $\gamma_{g}^{t} (gx) \leq 0$ since $gx \in gF$. But this is a contradiction as $gx = x$.\qed
\end{proof}

\begin{lemma}\label{lemma:fix_Ginvariant}
If $G$ is Abelian, then for every $g \in G$, the set $\fix(g)$ is $G$-invariant, i.e., $h \fix(g)=\fix(g)$ for all $h\in G$. 
\end{lemma}

\begin{proof}
Let $g, h \in G$. We show that $h \fix(g) = \fix(g)$. Indeed, if $y \in h \fix(g)$, i.e., $y = hx$ for some  $x \in \fix(g)$, then $gy = g (hx) = h(gx) = hx = y$. Therefore, $y \in \fix(g)$, and thus $h \fix(g) \subseteq \fix(g)$. The inclusion $\fix(g) \subseteq h \fix(g)$ follows by applying the previous argument to $h^{-1}$, implying that $h^{-1} \fix(g) \subseteq  \fix(g)$.\qed
\end{proof}

\begin{lemma}\label{lemma:fd_fixed_space}
Given a finite group $G \leq \Ort_{n}(\R)$, let $F \subseteq \R^{n}$ be a subgroup consistent fundamental domain for the collection $\{H_g\}_{g\in G}$, where $H_g=\{x \in \R^{n}:\gamma_g^tx\ge0\}$. %For a given $g\in G$, suppose that $F\cap H_g^=$ is a facet of~$F$. %Let $\gamma_{g} \in \R^{n}$ be a normal vector of the supporting hyperplane of $F$ at $F \cap gF$. 
Then $\gamma_{g}$ belongs to the orthogonal complement of the fixed space of $g$, i.e.,
\begin{equation*}
\gamma_{g} \in \fix(g)^{\perp} := \{x \in \R^{n} \talque gx = x \}^{\perp}.
\end{equation*}
\end{lemma}

\begin{proof}
We start by showing the lemma for the case that $G$ is Abelian. By Lemma~\ref{lemma:fix_Ginvariant} we have that $\fix(g)$ is $G$-invariant for every $g \in G$, and hence $h\fix(g)=\fix(g)$ for any $h\in G$. Therefore,  
\begin{equation*}
    \fix(g) = \fix(g)\cap \left(\bigcup_{h \in G} hF\right) =  \bigcup_{h \in G} (\fix(g)\cap hF)=\bigcup_{h \in G} h(\fix(g)\cap F).
\end{equation*} 
Let $\text{span}(S)$ denote the linear span of a set $S$. Notice that $\dim(\text{span}(\fix(g)\cap F))=\dim(\fix(g))$, otherwise, $\fix(g)$ would be contained in the union of finitely many subspaces of strictly smaller dimension, which is clearly a contradiction. Since ${F \cap \fix(g)\subseteq \fix(g)}$, we conclude that $\text{span}( F \cap \fix(g) ) = \fix(g)$. As by Lemma~\ref{lemma:fix_interF} we have that $F \cap \fix(g) \subseteq H_g^= $, this implies that $\fix(g) = \text{span}(F \cap \fix(g) )\subseteq H_g^=$. Since by definition $\gamma_g$ is orthogonal to every vector in $H_g^=$, we conclude that $\gamma_g \in \fix(g)^{\perp}$. The lemma follows if $G$ is Abelian.

For the general case, assume that $F$ is subgroup consistent for collection $\{H_h\}_{h\in G}$. Therefore, the Abelian subgroup $G'=\langle g\rangle$ has $F'=\bigcap_{h\in G'} H_h$ as a fundamental domain. Then our argument for the Abelian case implies that $\gamma_g \in \fix(g)^{\perp}$.\qed
\end{proof}

The following is the main contribution of this section.

\begin{theorem}\label{thm:FD_facet_subgroup_consistent}
Given a finite group $G \leq \Ort_{n}(\R)$, let $F \subseteq \R^{n}$ be a subgroup consistent fundamental domain for a collection $\{H_g\}_{g\in G}$, where $H_g=\{x:\gamma_g^tx\ge0\}$. Then, for every $g\in G$ there exists $\alpha_{g} \in \R^{n}$ such that $\gamma_{g} = (\id - g) \alpha_{g}.$
In particular, any facet-defining inequality for $F$ is of the form $\alpha_{g}^{t} x \geq \alpha_{g}^{t} g^{-1} x$ for some $g\in G$, and hence it is of Dirichlet type.
\end{theorem}

\begin{proof}
Recall that any automorphism $f$ of $\R^{n}$ satisfies $ \im(f)^{\perp} = \ker(f^t)$. Since ${\fix(g)=\ker(\id-g)}$, and recalling that $g^{-1}=g^t$ (interpreting $g$ as a matrix), by Lemma~\ref{lemma:fd_fixed_space} we have that
\begin{equation*}
    \gamma_{g} \in \fix(g)^\perp = \fix(g^{-1})^\perp = \ker(\id-g^t)^\perp = \im \left(\id-g \right).
\end{equation*}
Hence, there exists $\alpha_{g} \in \R^{n}$ such that $\gamma_{g} = (\id - g) \alpha_{g}.$ \qed
\end{proof}

\noindent{\textbf{Remark.}} It is worth noticing that this theorem does not imply that every subgroup consistent fundamental domain is a Dirichlet fundamental domain. The difference relays in the fact that in Dirichlet domains $\alpha=\alpha_g$ for all $g\in G$, while in subgroup consistent fundamental domains one can have different vectors $\alpha_g$ for different group elements $g$. For concrete examples see Section~\ref{sec:GDD}.

We say that a fundamental domain $F$ is  \emph{exact} if for every facet $S$ of $F$ there exists a group element $g\in G$ such that $S=F\cap gF$. 
In this case we say that $g$ defines a facet of $F$. Notice that it also holds that $S=F\cap H_g^=$. Exact fundamental domains are well structured and have been studied in the literature~\cite{ratcliffe19}. It is worth noticing that Dirichlet domains are exact.%, as well as all fundamental domains considered later in this article.

For exact fundamental domains, facets come in pairs, i.e., if $g$ defines a facet of $F$, then $g^{-1}$ also does. The proof of the following proposition can be found in Ratcliffe~\cite[Thm. 6.7.5.]{ratcliffe19}.

\begin{proposition}%{(Ratcliffe~\cite[Thm. 6.7.5.]{ratcliffe19})}
\label{prop:sidepairing}
Let $F \subseteq \R^{n}$ be an exact fundamental domain for $G \leq \Ort_{n}(\R)$ finite. If $S$ is a facet of $F$, then there is a unique non-trivial element $g \in G$ such that $S = F \cap g F$, moreover $g^{-1} S$ is a facet of $F$.
\end{proposition}
% \begin{proof}
% Let us consider an arbitrary $x \in F \cap g F$ and let $( x^{i} )_{i \in \N}$ be a sequence of points in the interior of $F$ that converges to $x$, which exists as $F$ is convex and has non-empty interior. Note that for every $i \in \N$, $\gamma^{t}x^{i} > 0$, then $g x^{i} \in \inte (gF)$ because $g$ is continuous and its inverse is also continuous. If we apply the orthogonal transformation $h$ we obtain that $hx^{i} \in h F$, $hgx^{i} \in \inte (hgF)$, and $hx \in h F \cap hg F$. Recall that by hypothesis $hF \cap hgF$ is a facet of $F$, then in particular $hgx \in F$. Given that for every $i \in \N$,  $(h \gamma)^{t} h gx^{i} = \gamma^{t} g x^{i} < 0$, and $h gx^{i} \to hgx$ as $i \to \infty$, i.e. $hgx$ is approximated by points from the open half-space where $F$ lies. Then, because $x$ and its sequence were arbitrary, $\inte (F) \cap \inte (hg F) \neq \emptyset$, but this is absurd because $F$ is a fundamental domain for $G$. Therefore, $hg = \id \iff h = g^{-1}$.\qed 
% \end{proof}

Proposition~\ref{prop:sidepairing} and Theorem~\ref{thm:FD_facet_subgroup_consistent} together imply the following corollary which gives a stronger connection between the facets $F\cap gF $ and $F\cap g^{-1}F$. Informally, the corollary says that we can take $\alpha_g = \alpha_{g^{-1}}$ in Theorem~\ref{thm:FD_facet_subgroup_consistent}. %that if $\alpha_{g}^{t} x \geq \alpha_{g}^{t} g x$ is facet-defining , then $\alpha_{g}^{t} x \geq \alpha_{g}^{t} g^{-1}x$ is also facet-defining. That is, not only do facets come in pairs, but we can also set $\alpha_{g^{-1}}=\alpha_{g}$. 

\begin{corollary}\label{cor:pairs}
Let $F \subseteq \R^{n}$ be an exact and subgroup consistent fundamental domain for $G \leq \Ort_{n}(\R)$ finite. Suppose that $H_g=\{x: \gamma_{g}^{t}x \ge 0\}$ defines the facet $F\cap gF=F\cap H_g^=$ and $H_{g^{-1}}=\{x: \gamma_{g^{-1}}^{t}x \ge 0\}$ defines the facet $F\cap g^{-1}F=F\cap H_{g^{-1}}^=$. %let $\alpha_{g}$ be a vector such that $\gamma_{g} = (\id - g) \alpha_{g}$´.
Then there exists a vector $\alpha_g$ such that 
\[
 H_g=\{x: \alpha_g^t x\ge \alpha_g^t (g^{-1}x)\} \text{ and } H_{g^{-1}}=\{x: \alpha_g^t x\ge \alpha_g^t (g x)\}.
\]

%the vector  $\gamma_{g^{-1}}=(\id - g^t) \alpha_{g}$ defines the facet $F\cap g^{-1}F$, and hence we can take $\alpha_{g^{-1}}=\alpha_{g}$.
% Then $\alpha_{g} = \alpha_{g^{-1}}$.
\end{corollary}
\begin{proof}
By Theorem~\ref{thm:FD_facet_subgroup_consistent}, $\gamma_{g} = (\id - g) \alpha_{g}$ for some $\alpha_g$ and hence $H_g=\{x: \alpha_g^t x\ge \alpha_g^t (g^{-1}x)\}$. Now, for any $x \in F \cap g^{-1}F$, we have $g x  \in g F \cap F=F\cap H_g^=$, and hence $\gamma_{g}^{t}(gx)=(g^{-1}\gamma_g)^tx=0$. Thus, $g^{-1}\gamma_g$ is orthogonal to $F \cap g^{-1}F$. As $\text{dim}(F \cap g^{-1}F)=n-1$, we obtain that $H_{g^{-1}}^= =\{x: (g^{-1}\gamma_g)^tx=0\}$. Now, notice that $H_{g^{-1}} =\{x: (g^{-1}\gamma_g)^tx\le0\}$. Indeed, if $H_{g^{-1}} =\{x: (g^{-1}\gamma_g)^tx\ge0\}$ we have that for any $x\in \inte(g^{-1}F)\neq \emptyset$ it holds that $(g^{-1}\gamma_g)^t x <0$, and hence $gx\in \inte(F)$ satisfies $\gamma_g^t(gx)<0$, which contradicts the construction of $H_g$. We conclude that $H_{g^{-1}} =\{x: (g^{-1}\gamma_g)^tx\le0\}$. The results follows by recalling that $\gamma_g = (\id - g) \alpha_{g}$, which implies that $H_{g^{-1}} = \{x: \alpha_g^t x\ge \alpha_g^t (g x)\}$.\qed

% by the hypothesis we have that 
% \begin{equation*}
%     \alpha_{g}^{t} x = \alpha_{g}^{t} gx \implies \alpha^{t}_{g} (\id - g) x = (\id - g^{-1}) \alpha_{g} x = 0
% \end{equation*}
% Therefore, $(\id - g^{-1})^{t} \alpha_{g}$ is a supporting hyperplane for the facet $g^{-1} F \cap F$. \qed
\end{proof}

\section{Generalized Dirichlet Domains}
\label{sec:GDD}

In this section we present our main contribution: an algorithm which constructs a fundamental domain for an arbitrary finite orthogonal group. We use the insights gained from the geometric properties of subgroup consistent and exact fundamental domains to guide our search for new constructions. In particular we create subgroup consistent fundamental domains based on a sequence of nested stabilizers of the $G$-action on~$\R^{n}$. This construction generalizes Dirichlet domains, and hence $k$-fundamental domains, as well as the \emph{Schreier-Sims fundamental domain}, presented in Section~\ref{subsection_ss}. Both types of fundamental domains can be easily constructed using our algorithm. Moreover, in Section~\ref{section_mul-uni} we exploit the flexibility of our construction to define a new fundamental domain with better properties for a specific group.

Theorem~\ref{thm:FD_facet_subgroup_consistent} and Corollary~\ref{cor:pairs} suggest that we should consider vectors $\alpha_g$ for some $g\in G$ and consider inequalities of the form $\alpha_g^tx\ge \alpha_g^t gx $ and $\alpha_g^tx\ge \alpha_g^t g^{-1}x $, although it seems hard to decide whether we should pick different vectors $\alpha_g$ for each pair $g, g^{-1}$, and if so, how to choose them. For instance, if we fix a vector $\alpha=\alpha_g$ for all $g\in G$, we would obtain a Dirichlet domain. However, if $\alpha$'s stabilizer is non trivial, then all inequalities $\alpha_g^tx\ge \alpha_g^t g^{-1}x $ in a coset of $G_{\alpha}$ are equivalent. This hints that we should choose a vector $\alpha$, apply a coset decomposition using a stabilizer subgroup, and add the Dirichlet inequalities related to all members of the decomposition. 

Furthermore, since all the elements of the group that fix $\alpha$ constitute a subgroup, if a fundamental domain $F$ for this subgroup were available, \emph{residual symmetries} could be taken care of with $F$, while \emph{non-residual symmetries} could be exploited via $\alpha^{t} x \geq \alpha^{t} (gx)$ for $g \notin G_{\alpha}$. Our next result points in this direction and lays the ground for our generalized Dirichlet domain algorithm.

\begin{theorem}\label{gdd_theorem}
Let $\alpha \in \R^{n}$ be an arbitrary vector and consider the polyhedral cone
\begin{equation*}
    F_{\alpha} = \{ x \in \R^{n} \talque \alpha^{t} x \geq \alpha^{t} gx, \squad \forall g \in G\}.
\end{equation*}
Suppose that $F$ is a fundamental domain for the subgroup $G_{\alpha}$, i.e., the pointwise stabilizer of $\alpha$. Then $F \cap F_{\alpha}$ is a fundamental domain for $G$.

Moreover, for any transversal $T$ for $G_{\alpha}$ in $G$, the polyhedral cone $F_{\alpha}$ can be described as
\begin{align*}
    F_{\alpha} = \{ x \in \R^{n} \talque \alpha^{t} x \geq \alpha^{t} gx, \squad \forall g \in T \cup T^{-1} \},
\end{align*}
where $T^{-1}:= \{ g^{-1}  \talque g \in T\}.$
\end{theorem}
\begin{proof}
First, notice that $F \cap F_{\alpha}$ is closed and convex. Now, we show that every $x \in \R^{n}$ has a representative in $F \cap F_{\alpha}$. In other words, we show that there exists some $g \in G$ such that $gx \in F \cap F_{\alpha}$. Let us consider two cases: (i) $x \in F_{\alpha}$ and (ii) $x \notin F_{\alpha}$. In case (i), since $F$ is a fundamental domain for $G_{\alpha}$, there exists $g \in G_{\alpha}$ such that $gx \in F$. As $ \alpha = g^{-1} \alpha$ we have that
\begin{align*}
    \alpha^{t}gx = (g^{t} \alpha)^{t} x = (g^{-1}\alpha)^{t} x = \alpha^{t} x.
\end{align*}
Therefore, $gx \in F \cap F_{\alpha}$. Now, consider case (ii), i.e., there exists $g' \in G$ such that $\alpha^{t} g' x > \alpha^{t} x$. Let $h \in \argmax \{ \alpha^{t} g x \talque g \in G \}$. Clearly $hx\in F_{\alpha}$, and hence we are done if $hx \in F$. If $hx \not\in F$ there exists $\tilde{g} \in G_{\alpha}$ such that $\tilde{g}(hx) \in F$. We conclude that $\tilde{g}(hx) \in F\cap F_{\alpha}$ by the same argument as in case (i).

Now we prove that for any $g \in G\setminus\{\id\}$ we have that $\text{int} (F \cap F_{\alpha})$ and $\text{int} (g(F \cap F_{\alpha}))$ are disjoint. Let $x \in \interior (F \cap F_{\alpha})$. It suffices to show that $gx \notin F \cap F_{\alpha} $. Indeed, since $x \in \interior(F) \cap \interior (F_{\alpha})$, then $gx \notin F \cap F_{\alpha}$ for all $g \in G_{\alpha} \setminus \{ \id \}$ as $F$ is a fundamental domain for $G_{\alpha}$. Moreover, since $x$ belongs to the interior of $F_{\alpha}$, it holds that $\alpha^{t} x > \alpha^{t} gx$ for all $g \in G \setminus G_{\alpha}$. Therefore, $gx \notin F \cap F_{\alpha}$. We conclude that $F\cap F_{\alpha}$ is a fundamental domain for $G$.

Let us show that it suffices to consider the Dirichlet type inequalities associated to a transversal and its inverses $T\cup T^{-1}$ to describe $F_{\alpha}$. Let $T \subseteq G$ be a transversal for $G_{\alpha}$ and let $g \in G$. If $g \in G_{\alpha}$, then clearly $\alpha^t = \alpha^t g$ and hence the inequality $\alpha^tx = \alpha^t gx$ is trivial. Consider $g \notin G_{\alpha}$, and thus $g^{-1} \notin G_{\alpha}$. Then there exists $r \in T$ such that $g^{-1} \in r G_{\alpha}$, i.e., $g^{-1} \alpha = r \alpha$. Therefore,
\begin{equation*}
    \alpha^{t} (gx) = (g^{-1} \alpha)^{t} x = (r\alpha)^{t} x = \alpha^{t} (r^{-1}x),
\end{equation*}
from which we can conclude that $\alpha^{t} x \geq \alpha^{t} (gx)$ and $\alpha^{t} x \geq \alpha^{t} (r^{-1}x)$ are equivalent.\qed
\end{proof}

%Also it is hard to decide if we should pick a different vector $\alpha_g$ for each $g,g^{-1}$ pair, and if so how to choose them. On the other hand, if we fix a vector $\alpha=\alpha_g$ for all $g\in G$, it seems we would obtain again a Dirichlet domain. However, there is one exception, which is the case that $\alpha$ has not a trivial stabilizer. Indeed, if $G_{\alpha}$ is not trivial, all inequalities $\alpha_g^tx\ge \alpha_g^t g^{-1}x $ in a coset of $G_{\alpha}$ are equivalent. This suggests to choose a vector $\alpha$ and to iteratively apply coset decompositions using stabilizer subgroups, and adding Dirichlet inequalities for each member in the decomposition. To this end, we first identify a vector $\alpha_{1}$ in $\R^{n}$ that is not fixed by $G$ and compute its stabilizer $G_{1}$. Then we pick a transversal $H$ for $G_{1}$, that is, a collection of representatives of the (left) cosets of $G_{1}$ in $G$, so that $G = \bigcup_{h\in H} h G_{1}.$

%For each $h \in H \cup H^{-1}$, we add its corresponding Dirichlet inequality and thus we separate points in $\R^{n}$ which are moved by at least one coset representative but are fixed by $G_{1}$. Then we recurse on $G_{1}$ to separate points which are not fixed by $G_{1}$. See \ref{algo:GeneralizedDirichletDomain} for the precise description. %Our fundamental domain derives its name from the fact that in each step we compute a Dirichlet domain with center $\alpha_{i}$ for a suitable coset decomposition.

An iterative application of Theorem~\ref{gdd_theorem} yields Algorithm~\ref{algo:GeneralizedDirichletDomain}. We say that a fundamental domain constructed by this algorithm is a \emph{generalized Dirichlet domain} (GDD). See Examples~\ref{ex:non_exact_permutation} and~\ref{ex:notexact} below for concrete examples of this construction.

\begin{algorithm}[H]
\caption{Construction of a generalized Dirichlet domain (GDD)}\label{algo:GeneralizedDirichletDomain}
\begin{algorithmic}[1]
\Statex \textbf{Input}: A set of generators $S_{G}$ of a finite orthogonal group $G$
\Statex \textbf{Output}: A fundamental domain $F$ for $G$
% \Statex
% \Procedure{algo:GeneralizedDirichletDomain}{$S_{G}$}
\State Set $F := \R^{n}$,  $G_{0} := G$, and $i:= 1$
\While{$G_{i-1}\neq \{\id\}$}\label{algo_while}
    \State\label{algo_line6}Choose $\alpha_{i} \in \R^{n}$ such that $g \alpha_{i} \neq \alpha_{i}$ for some $g \in G_{i-1}$
    \State\label{algo_compute}Compute $G_{i}:= \{g \in G_{i-1} \talque g \alpha_{i} = \alpha_{i} \}$
    \State\label{algo_choose}Choose a transversal $T_{i}$ for $G_{i}$ in $G_{i-1}$ and add the inverses $R_{i} := T_{i} \cup T_{i}^{-1}$
    \State Set $F_{i} := \{x \in \R^{n} \talque \alpha_{i}^{t} x \geq \alpha_{i}^{t} h x \quad \forall h \in R_{i} \}$
    \State $F:= F \cap F_{i} $ and $i:=i+1$
\EndWhile
\State \textbf{return} $F$
% \EndProcedure
\end{algorithmic}
\end{algorithm}

\begin{theorem}\label{thm:Gdd_Algo}
%Given a set of generators for a finite orthogonal group $G \leq \Ort_{n}(\R)$, 
Algorithm~\ref{algo:GeneralizedDirichletDomain} terminates in at most $n$ iterations and outputs a %an exact 
fundamental domain $F$.
\end{theorem}
\begin{proof}
We follow an inductive bottom-up argument. First we prove the base case. Given the output of Algorithm~\ref{algo:GeneralizedDirichletDomain} let $m$ be the smallest integer such that $G_{m} = \langle \id \rangle$. Notice that $m \leq n$ since the set $\{ \alpha_{i} \}_{i=1}^{m}$ must be linearly independent, otherwise some $\alpha_{i}$ would belong to the linear span of $\{ \alpha_{j} \}_{j<i}$ implying that $\alpha_{i}$ is fixed by $G_{i-1}$, which is a contradiction. Therefore the algorithm terminates in at most $n$ iterations. Then, the transversal $T_{m}$ for $G_{m}$ in $G_{m-1}$ computed in Line~\ref{algo_choose} corresponds to $G_{m-1}$, i.e. $G_{m-1}$ is trivially decomposed by $G_{m}$. Hence, $F_{m} = F_{\alpha_{m}}$ is a Dirichlet domain for $G_{m-1}$. Therefore, $F_{m-1} \cap F_{m}$ is a fundamental domain for $G_{m-2}$ by Theorem~\ref{gdd_theorem}. Consequently, $F$ is a fundamental domain for $G$ since $\cap_{i=2}^{m} F_{i}$ is a fundamental domain for $G_{1}$, by iteratively applying Theorem~\ref{gdd_theorem}.\qed
\end{proof}

It is worth noticing that if we take $\alpha_1$ such that $G_1$ is trivial, then the algorithm finishes after one iteration. Indeed, the obtained fundamental domain is the Dirichlet domain $F_{\alpha_1}$. This justifies the name generalized Dirichlet domain.

We already know that  Algorithm~\ref{algo:GeneralizedDirichletDomain} terminates after at most $n$ iterations. For the rest of the analysis of the running time, we focus on permutation groups. Lines \ref{algo_compute} and \ref{algo_choose} are the most challenging with respect to the algorithm's computational complexity. 
The result of the computation in line \ref{algo_compute} is a setwise stabilizer of the coordinates of $\alpha$. Computing a set of generators for this subgroup can be performed in quasi-polynomial time with the breakthrough result by Babai~\cite{babai2016grapharx,babai2016graphSTOC} for String Isomorphism. In general, however, $R_1$ might be of exponential size. Indeed, the number of cosets of $G_1$ in $G$ equals the size of the orbit $|\text{Orb}_G(\alpha_1)|$, by the Orbit-Stabilizer Theorem. If we choose $\alpha_1$ with pairwise different coordinates, then $|\text{Orb}_G(\alpha_1)|=|G|$. This is exactly the case for the Dirichlet domain. 

On the other hand, by choosing the $\alpha_i$ vectors carefully we can avoid the described problem. In particular, suppose that $\alpha = (\alpha^{(1)}, \dots, \alpha^{(k)}, 0, \dots, 0)$ such that $\alpha^{(i)} \neq \alpha^{(j)}$ for $i \neq j$ in $[k]$, and $\alpha^{(i)} \neq 0$ for $i \in [k]$. Hence, a set of generators for the stabilizer $G_\alpha$ can be computed in polynomial time. Indeed, it corresponds to the pointwise stabilizers of coordinates $1$ to $k$~\cite[Section 5.1.1]{seress03}. Moreover, the number of cosets is $O(n^k)$, as again the number of cosets equals the cardinality of the orbit of $\alpha$. In other words, we have just proven the following proposition.
\begin{proposition}
Let $G \leq S_{n}$ a permutation group and let $k \in [n]$ be a constant. Suppose that each $\alpha_{i}$ in Algorithm~\ref{algo:GeneralizedDirichletDomain} satisfies
\begin{equation*}
    \alpha_{i} = (\alpha_{i}^{(1)}, \dots, \alpha_{i}^{(k)}, 0, \dots, 0),
\end{equation*} 
$\alpha_{i}^{(\ell)} \neq \alpha_{i}^{(m)}$ for $\ell \neq m$ in $[k]$, and $\alpha_{i}^{(\ell)} \neq 0$ for $i \in [k]$. Then the associated GDD for $G$ can be computed in time $O(n^{O(k)})$.
\end{proposition}

\vspace{0.3em}

\subsection{Geometric Properties of Generalized Dirichlet Domains}

In this section we study two important geometric properties of fundamental domains: subgroup consistency and exactness. To this end, first we show how can a \emph{canonical} representation for generalized Dirichlet domains be defined via a partition of $G$ into \emph{layers}. Then, we use this representation for GDDs to show that they are subgroup consistent. Moreover, we show that they are not necessarily exact.

\vspace{1em}

\noindent\textbf{Subgroup consistency.} Let $F = \bigcap_{i=1}^{m} F_{i}$ be the output of Algorithm~\ref{algo:GeneralizedDirichletDomain}, where $m \in [n]$ denotes the smallest index such that $G_{m} = \langle \id \rangle$, and
\begin{equation*}
    F_{i} = \{x \in \R^{n} \talque \alpha_{i}^{t} x \geq \alpha_{i}^{t} h x \quad \forall h \in R_{i} \}
\end{equation*}
where $R_{i} = T_{i} \cup T_{i}^{-1}$, with $T_{i}$ a transversal for $G_{i}$ in $G_{i-1}$ and $G_{0} := G$. We will define a partition of $G$ and a representation of $F$ using the nested coset decompositions produced by Algorithm \ref{algo:GeneralizedDirichletDomain}. Notice that in the $i$-th iteration of our GDD algorithm a coset decomposition is computed using the subgroup $G_{i}$. In other words, $G_{i-1}$ is partitioned into cosets as
\begin{equation*}
    G_{i-1} = \bigcup_{g \in T_{i}} g G_{i}
\end{equation*}
Now, fix $i \in [m]$. We say that $g \in G \setminus \{ \id \}$ belongs to the \emph{$i$-th layer $L_{i}$ of $G$ induced by $\{\alpha_{i} \}_{i=1}^{m}$} if $g \in G_{i-1}\setminus G_{i}$. Since $G_{j-1} \geq G_{j}$ for all $j \in [m]$, $g$ belongs to $L_{i}$ if and only if $i$ smallest index such that $g \in G_{i}$. Therefore, letting $\id \in L_{m}$, we have that $\{L_{i} \}_{i=1}^{m}$ is a partition of $G$ as every $g$ belongs to a unique $L_{j}$.

Now we are ready to define a \emph{GDD representation} of $F$. First, note that $T_{i} \setminus \{ \id \} \subseteq L_{i}$ since by definition $T_{i} \subseteq G_{i-1}$ and $T_{i} \setminus \{ \id \} \cap G_{i} = \emptyset$. The latter implies that $T^{-1}_{i} \setminus \{ \id \} \subseteq L_{i}$, hence $R_{i} \setminus \{ \id \} \subseteq L_{i}$. Now, let $g \in G \setminus \{ \id \}$. Then $g \in L_{j}$ for some $j \in [m]$. Since $g \notin G_{j}$, then so does $g^{-1}$, and it belongs to some coset $r G_{j}$, where $r \in T_{j}$.  Since $g^{-1} \alpha_{j} = r \alpha_{j}$, then 
\[
     \alpha_{j}^{t}gx = (g^{-1} \alpha_{j})^{t}x 
     = (r \alpha_{j})^{t} x 
     = \alpha_{j}^{t} r^{-1} x.
\]
Since $r^{-1}$ induces the same Dirichlet type inequality as $g$ we say that $g$ is \emph{associated to} $r^{-1}$. Therefore, any $g \in G$ can be associated to some $h \in R_{j}$ in some layer $L_{j}$ of $G$, i.e., we can define
\begin{equation*}
    H_{g} := \{ x \in \R^{n} \talque \alpha_{j}^{t} x \geq \alpha^{t}_{j} hx \},
\end{equation*}
and say that $\{ H_{g} \}_{g \in G}$ is a \emph{GDD representation of $F$ induced by $\{ \alpha_{i} \}_{i=1}^{m}$}. 

% \begin{remark}
% Notice that, given $\{ \alpha_{i} \}_{i=1}^{m}$, the GDD canonical representation of $F$ is unique. Moreover, we have a unique $\alpha \in \R^{n}$ per layer of $G$. For instance, given that the construction of the Dirichlet domain produces only one layer for $G$, there is only one $\alpha$ for all group elements.
% \end{remark}

\begin{proposition}
Generalized Dirichlet domains are subgroup consistent.
\end{proposition}
\begin{proof}
Let $\{ H_{g} \}_{g \in G}$ be the \emph{GDD representation} of $F$ induced $\{ \alpha_{i} \}_{i=1}^{m}$ obtained via Algorithm~\ref{algo:GeneralizedDirichletDomain}. Let $G'$ be any subgroup of $G$. We show that $F' := \bigcap_{g \in G'} H_{g}$ is a fundamental domain for $G'$.

Clearly $F'$ is closed and convex. Let us prove that for all $x \in \R^{n}$, there exists some $g \in G'$ such that $gx \in F'$. Suppose that $x \notin F'$ and let $i_{1} \in [m]$ denote the first layer such that $\alpha_{i_{1}}^{t} x < \alpha_{i_{1}}^{t} gx$ for some $g \in L_{i_{1}} \cap G'$. Let $h_{i_{1}} \in \argmax \squad \{ \alpha_{i_{1}}^{t} g x \talque g \in L_{i_{1}} \cap G'\}$. Notice that $\alpha_{j}^{t} x = \alpha_{j}^{t} h_{i_{1}} x$ for all $j < i_{1}$ since $h_{i_{1}} \in G_{j}$ for $j < i_{1}$, and then 
\begin{equation*}
    h_{i_{1}} x \in \bigcap_{ \substack{i \in [i_{1}] : \\ g \in L_{i} \cap G'}} H_{g}.
\end{equation*}
If $\tilde{x} := h_{i_{1}} x \notin F'$, we can replicate the argument on the first layer $i_{2} > i_{1}$ such that $\tilde{x} \notin H_{\tilde{g}}$ where $\tilde{g} \in L_{i_{2}} \cap G'$. Inductively we obtain an element $h := h_{i_{\ell}} h_{i_{\ell-1}} \cdots h_{i_{1}} \in G'$ such that $h x \in F'$.

Now, we prove that $\interior (F') \cap g \interior (F') = \emptyset$ for any non-trivial $g \in G'$. It suffices to show that if $x \in \interior (F')$ then $g x \notin F'$. Indeed, since $x \in \interior (F')$, we have that
\begin{align*}
    &\alpha_{i}^{t} x > \alpha_{i}^{t} r x \hspace{1em} \text{for all $i \in [m]$ and $r \in L_{i} \cap G'$} \\
    \iff &\alpha_{i}^{t} x > \alpha_{i}^{t} h x \hspace{1em} \text{for all $i \in [m]$ and $h \in R_{i}$ associated to some $r \in G'$.}
\end{align*}
Suppose $g$ belongs to layer $L_{j}$ and it is associated to some $h \in H_{j}$. Therefore, $\alpha_{j}^{t} gx = \alpha_{j}^{t} h$ and 
\begin{align*}
    \alpha_{j}^{t} gx &< \alpha_{j}^{t} x = \alpha_{j}^{t} g^{-1} (g x)
\end{align*}
where $g^{-1} \in L_{j} \cap G'$, i.e., $gx \notin F'$.\qed
\end{proof}

\color{black}

\vspace{1em}

\noindent\textbf{Exactness.} Recall that a fundamental domain is exact if every facet $S$ of $F$ is of the form $S = F \cap gF$ for some $g \in G$. Dirichlet domains are exact~\cite[Theorem 6.7.4]{ratcliffe19} though generalized Dirichlet domains may not be exact. This means that for some iteration $i$ of Algorithm~\ref{algo:GeneralizedDirichletDomain} there exists some $h \in R_{i}$ such that $F \cap H_{h}^{=}$ is a facet of $F$ and it satisfies:
    \begin{equation*}
        \relint (F \cap H^{=}_{h}) \cap hF \neq \emptyset \hspace{0.5em} \text{and} \hspace{0.5em}  \relint (F \cap H^{=}_{h}) \cap gF \neq \emptyset
    \end{equation*}
for some $g \neq h$. The following examples show non-exact GDDs. The first one is an example for a permutation group in $\mathbb{R}^4$. The second one is a more geometrical example in $\mathbb{R}^3$.

\begin{example}\label{ex:non_exact_permutation}
Let $g := (1 \: 2 \: 3 \: 4)$  and consider $G := \langle g \rangle$. We construct a GDD for $G$ with $\alpha_{1} := (1,0,1,0)$ and $\alpha_{2} := ( 1,0,0,0)$.

Indeed, if we first choose $\alpha_{1}$, its stabilizer is $G_{\alpha_{1}} = \{ \id, g^{2} \}$, and a $G_{\alpha_{1}}$-transversal is $H_{1} = \{ \id, g \}$. Hence, 
\begin{equation*}
    F_{1} = \{ x \in \R^{4} \talque x_{1} + x_{3} \geq x_{2} + x_{4} \}.
\end{equation*}
Next, the only subgroup of $G_{\alpha_{1}}$ that stabilizes $\alpha_{2}$ is $\langle \id \rangle$, hence
\begin{equation*}
    F_{2}= \{ x \in \R^{4} \talque x_{1} \geq x_{3} \}.
\end{equation*}
The resulting GDD for $G$ is
\begin{equation*}
    F := F_{1} \cap F_{2} = \{ x \in \R^{4} \talque x_{1} + x_{3} \geq x_{2} + x_{4}, \squad x_{1} \geq x_{3} \}.
\end{equation*}
Now, we exhibit two points that certify the non-exactness of $F$. Consider $x = (2,2,1,1)$ and $\tilde{x} = (2,1,1,2)$. Clearly both points belong to the facet $F \cap H_{g}^{=}$, and since none of them satisfy $x_{1} = x_{3}$ they belong to the relative interior of $F \cap H_{g}^{=}$. Moreover, as $g^{-1}x = (2,1,1,2) \in F \iff x \in gF$ and $g\tilde{x} = (2,2,1,1) \in F \iff \tilde{x} \in g^{-1}F$, then the relative interior of a facet of $F$ intersects $gF$ and $g^{-1}F$. Therefore $F$ is not exact. \xqed

\end{example}

\begin{example}\label{ex:notexact}
Let us consider the three-dimensional space $\R^3$, and let $g$ be the isometry that consists of a rotation by $90$ degrees around the $x_{3}$-axis, followed by a reflection with respect to the plane $\text{span}(e_1,e_2)$. The matrix associated to $g$ is 
\begin{equation*}
    \begin{pmatrix}
        0 & -1 & 0 \\
        1 & 0 & 0 \\
        0 & 0 & -1
    \end{pmatrix}.
\end{equation*}
The group $G=\langle g \rangle = \left\{ \id, g, g^2, g^3  \right\}$ comprises four different elements. By taking $\alpha_{1} = (0,0,1)$, and $\alpha_{2} = (0,1,0)$ when running Algorithm~\ref{algo:GeneralizedDirichletDomain}, where $G_{\alpha_{1}} = \{ \id, g^{2} \}$, we obtain the following GDD
\begin{equation*}
    F = \left\{ x \in \R^3 \talque x_2 \geq 0, x_3 \geq 0 \right\}.
\end{equation*}
Then, we have
\begin{align*}
    gF &= \left\{ x\in\R^3 \talque x_1 \leq 0, x_3 \leq 0 \right\}, \\
    g^2F &= \left\{ x\in\R^3 \talque x_2 \leq 0, x_3 \geq 0 \right\}, \\
    g^3F &= \left\{ x\in\R^3 \talque x_1 \geq 0, x_3 \leq 0 \right\}.
\end{align*} 
It is easy to see that this tessellation is not exact, see Figure~\ref{fig:notexact}.%of the space is not exact, as the line $\{ x\in \R^3 \talque x_2 = 1, x_3 = 0 \}$ belongs to the relative interior of $\partial F$ and comprises the point $(1, 1, 0)$ and $(-1, 1, 0)$, which belong to $gF$ and $g^3F$, respectively. An illustration of the situation is given in Figure~\ref{fig:notexact}.
\begin{figure}
\centering
\begin{tikzpicture} % Schéma général
% Coordinates
	\coordinate (origin) at (0,0);
	\node[below] (x) at (-1.2,-1.2) {$x_1$};
	\node[right] (y) at (2,0) {$x_2$};
	\node[above] (z) at (0,2) {$x_3$};

% gF	
	\draw[gfd] ($.5*(y)$) -- ($.5*(y)-.5*(x)$) -- ($-.5*(y)-.5*(x)$) -- ($-.5*(y)$) -- cycle ; % lower rectangle
	\draw[gfd] ($.5*(y)-.5*(z)$) -- ($.5*(y)-.5*(x)-.5*(z)$) -- ($-.5*(y)-.5*(x)-.5*(z)$) -- ($-.5*(y)-.5*(z)$) -- cycle ; % upper rectangle
	\draw[gfd] ($.5*(y)$) -- ($.5*(y)-.5*(z)$); 
	\draw[gfd] ($.5*(y)-.5*(x)$) -- ($.5*(y)-.5*(x)-.5*(z)$); 
	\draw[gfd] ($-.5*(y)-.5*(x)$) -- ($-.5*(y)-.5*(x)-.5*(z)$); 
	\draw[gfd] ($-.5*(y)$) -- ($-.5*(y)-.5*(z)$);
% g3F	
	\draw[g3fd] ($.5*(y)$) -- ($.5*(y)+.5*(x)$) -- ($-.5*(y)+.5*(x)$) -- ($-.5*(y)$) -- cycle ; % lower rectangle
	\draw[g3fd] ($.5*(y)-.5*(z)$) -- ($.5*(y)+.5*(x)-.5*(z)$) -- ($-.5*(y)+.5*(x)-.5*(z)$) -- ($-.5*(y)-.5*(z)$) -- cycle ; % upper rectangle
	\draw[g3fd, fill=cyan!50] ($.5*(y)$) -- ($.5*(y)-.5*(z)$) -- ($-.5*(y)-.5*(z)$) -- ($-.5*(y)$) -- cycle;
	%\draw[g3fd] ($.5*(y)$) -- ($.5*(y)-.5*(z)$); 
	\draw[g3fd] ($.5*(y)+.5*(x)$) -- ($.5*(y)+.5*(x)-.5*(z)$); 
	\draw[g3fd] ($-.5*(y)+.5*(x)$) -- ($-.5*(y)+.5*(x)-.5*(z)$); 
	%\draw[g3fd] ($-.5*(y)$) -- ($-.5*(y)-.5*(z)$);
% g2F	
	\draw[g2fd, fill=green!50] ($.5*(x)$) -- ($.5*(x)-.5*(y)$) -- ($-.5*(x)-.5*(y)$) -- ($-.5*(x)$) -- cycle ; % lower rectangle
	\draw[g2fd] ($.5*(x)+.5*(z)$) -- ($.5*(x)-.5*(y)+.5*(z)$) -- ($-.5*(x)-.5*(y)+.5*(z)$) -- ($-.5*(x)+.5*(z)$) -- cycle ; % upper rectangle
	\draw[g2fd] ($.5*(x)$) -- ($.5*(x)+.5*(z)$); 
	\draw[g2fd] ($.5*(x)-.5*(y)$) -- ($.5*(x)-.5*(y)+.5*(z)$); 
	\draw[g2fd] ($-.5*(x)-.5*(y)$) -- ($-.5*(x)-.5*(y)+.5*(z)$); 
	\draw[g2fd] ($-.5*(x)$) -- ($-.5*(x)+.5*(z)$); 	
% F	
	\draw[fd, fill=red!50] ($.5*(x)$) -- ($.5*(x)+.5*(y)$) -- ($-.5*(x)+.5*(y)$) -- ($-.5*(x)$) -- cycle ; % lower rectangle
	\draw[fd, fill=red!50] ($.5*(x)$) -- ($.5*(x)+.5*(z)$) -- ($-.5*(x)+.5*(z)$) -- ($-.5*(x)$) -- cycle;
	\draw[fd] ($.5*(x)+.5*(z)$) -- ($.5*(x)+.5*(y)+.5*(z)$) -- ($-.5*(x)+.5*(y)+.5*(z)$) -- ($-.5*(x)+.5*(z)$) -- cycle ; % upper rectangle
	%\draw[fd] ($.5*(x)$) -- ($.5*(x)+.5*(z)$); 
	%\draw[fd] ($-.5*(x)$) -- ($-.5*(x)+.5*(z)$); 
	\draw[fd] ($.5*(x)+.5*(y)$) -- ($.5*(x)+.5*(y)+.5*(z)$); 
	\draw[fd] ($-.5*(x)+.5*(y)$) -- ($-.5*(x)+.5*(y)+.5*(z)$); 
% Axes
	\draw[axes] ($-.8*(x)$) -- (origin) -- (x); 
	\draw[axes2] ($-.8*(y)$) -- ($-.27*(y)$);
	%\draw[axes2] ($-.27*(y)$) -- (origin);
	\draw[axes] (origin) -- (y); 
	\draw[axes2] ($-.8*(z)$) -- ($-.32*(z)$);
	%\draw[thick, gray!50] ($-.32*(z)$) -- (origin);
	\draw[axes] (origin) -- (z);
% Corrections 3D-view
	\draw[fd] ($.5*(x)$) -- ($.5*(x)+.5*(y)$) -- ($.5*(x)+.5*(y)+.5*(z)$)-- ($.5*(x)+.5*(z)$) -- cycle;
	\draw[fd] ($.5*(x)+.5*(z)$) -- ($-.5*(x)+.5*(z)$);
	\draw[g2fd] ($.5*(x)-.5*(y)$) -- ($.5*(x)-.5*(y)+.5*(z)$);
\end{tikzpicture}
\caption{The situation of Example~\ref{ex:notexact}, restricted to a cube. $F$ is the red domain, $gF$ the blue one, $g^2F$ the green one, and $g^3F$ the cyan one.}
\label{fig:notexact}
\end{figure}
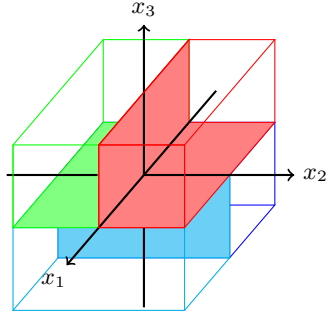\xqed
\end{example}

\subsection{The Lex-Max Fundamental Domain}%\label{subsection_ss}

In this section, we study a natural idea for breaking symmetries: in any orbit, choose the vector that is lexicographically maximal. We relate this idea to our generalized Dirichlet domain construction. %We will show that the closure of the defined set is a Generalized Dirichlet Domain, which coincides with the Schreir-Sims inequalities introduced by Liberti and Ostrowskit~\cite{} and Salvagning~\cite{salvagnin18}. 
%present a convex fundamental domain with at most $n-1$ facets for an arbitrary permutation group $G \leq S_{n}$. The inequalities that define this polyhedron were introduced in \cite{salvagnin18,liberti+ostrowski14}, but it was not noted, nor proven, that this object is in fact a fundamental domain and that its number of defining inequalities is linear in $n$. To achieve this, we work with the set of \emph{lexicographically maximal vectors} under the $G$-action, study some of its structural properties, and characterize its closure with \emph{directed trees}. 

% In this section, we present a fundamental domain with at most $n-k$ facets for an arbitrary permutation group $G \leq S_{n}$ with $k$ $G$-orbits. The inequalities that define this polyhedron were introduced in \cite{salvagnin18,liberti+ostrowski14}, but it was not noted, nor proven, that this object is in fact a fundamental domain and that its number of defining inequalities is linear in the group's degree. To achieve this, we work with the set of \emph{lexicographically maximal vectors} under the $G$-action, study some of its structural properties, and characterize its closure with \emph{directed trees}. 

Let $\succ$ denote a lexicographic order on $\R^{n}$, that is for any pair $x,y \in \R^{n}$ we say that $y \succ x$ if there exists $j \in [n]$ such that $y_{j} > x_{j}$ and $y_{i} = x_{i}$ for all $i<j$. Therefore $\succeq$ defines a total order on $\R^{n}$, where, $y \succeq x$ if and only if $y \succ x$ or $y =x$.
% Given this relation, we define the total order $\succeq$ on $\R^{n}$ as: for any pair $x,y \in \R^{n}$,
%   \begin{equation*}
% y \succeq x \iff (y \succ x) \quad \text{or} \quad (y =x).
% \end{equation*}
Given a group $G \leq S_{n}$ acting on $\R^{n}$, we define
\begin{equation*}
    \Lex_{G} := \{ x \in \R^{n} \talque x \succeq gx,\,\forall g \in G \}.
\end{equation*}

In what follows we show an alternative characterization of the set of lexicographically maximal points using $k$-fundamental domains. Recall that a $k$-funda\-mental domains is a Dirichlet domain $F_{\overline{k}}$ where $\overline{k} = (k^{n-1}, k^{n-2}, \dots, k, 1)$ for some integer~$k \geq 2$.

\begin{lemma}\label{lemma:Monotonicity-Lex}
Let $n \in \N$ and $x, y \in \R^{n}$. If $x \lexx y$, then there exists $N \in \N$ such that for every integer $k> N$ it holds that $\overline{k}^{t} x > \overline{k}^{t} y$.%, where $\overline{k} := (k^{n-1}, k^{n-2}, \dots, k, 1)$ for $k \geq 2$ integer.
\end{lemma}
\begin{proof}
Suppose $x \lexx y$. This implies that $x_{1} > y_{1}$ or there exists $i \in [n] \setminus \{1 \}$ such that $x_{i} > y_{i}$ and $x_{j} = y_{j}$ for all $j \in [i-1]$. Let $c := x_{i} - y_{i} > 0$ such that $i$ is the smallest $i \in [n]$ for which $x_{i} - y_{i} >0$, and let $m := \max \{ | y_{j} - x_{j} | \talque j \in [n] \}$. Note that if $i = n$ our claim is trivially true since for any $k \in \N$ we have that $\overline{k}^{t} x - \overline{k}^{t} y = x_{n} - y_{n} >0$. If $i = n-1$ we have $\overline{k}^{t} x - \overline{k}^{t} y = kc + (x_{n} - y_{n})$, hence there exists $N \in \N$ such that $kc + (x_{n} - y_{n}) >0$ for all $k \geq N$. Now, suppose $i \in [n-2]$. Then, for $k \in \N$,
\begin{align*}
\overline{k}^{t} x - \overline{k}^{t} y% &= %(k^{n-1} x_{1} + \dots + k^{n-i-1} x_{i} + \dots + x_{n}) + \\
%& \qquad \qquad \qquad \qquad \qquad \qquad \quad- ( k^{n-1} y_{1} + \dots + k^{n-i-1} y_{i} + \dots + y_{n} ) \\
&= k^{n-i} c - k^{n-i-1}(y_{i+1} - x_{i+1})  \dots - k(y_{n-1} - x_{n-1}) - (y_{n} - x_{n}) \\
&\geq k^{n-i} c - m \sum_{j=0}^{n-i-1} k^{j} = k^{n-i} c - m \left( \frac{k^{n-i} - 1}{k-1} \right).
\end{align*}
 Now, note that $\lim_{k \to \infty} \left[ k^{n-i} c - m\left( \frac{k^{n-i} - 1}{k-1} \right) \right] = + \infty$, because the sign of the leading coefficient of this rational function is positive and the degree of the numerator is greater than that of the denominator. Hence, there exists $N$ such that $ k^{n-i} c - m\left( \frac{k^{n-i} - 1}{k-1} \right) >0$ for all $k>N$.% $\overline{k}^{t} x - \overline{k}^{t} y$ for all $$
%Then, there exists $k_{0} \in \N$ such that
% \begin{equation*}
% c > m \left( \frac{k_{0}^{n- i} - 1}{k_{0}^{n - i }(k_{0}-1)} \right).
% \end{equation*}
% Then it follows that $\overline{k_0}^{t} x > \overline{k_0}^{t} y$.

% Finally, let $\psi: \R_{+} \to \R$ be the rational function defined by $\psi(k) := m\left( \frac{k^{n- i} - 1}{k^{n - i }(k-1)} \right)$. Given that $\psi$ is rational and $\lim\limits_{k \to \infty} \psi(k) = 0$, $g$ is monotone decreasing from a suitable $k' \in \N$ onwards. \qed
\end{proof}

With the help of the previous lemma, we provide an alternative characterization of $\Lex_G$. Recall that a \emph{fundamental set} is a set that contains exactly one representative for each $G$-orbit.

\begin{lemma}\label{lemma:ss_liminf_lex}
Let $G \leq S_{n}$. Then $\Lex_{G}$ is a convex fundamental set and
\begin{equation*}
     \Lex_{G} = \bigcup_{i=1}^{\infty} \bigcap_{k=i}^{\infty} F_{\overline{k}}=\liminf_{k \to \infty} F_{\overline{k}}.
\end{equation*}
\end{lemma}

\begin{proof}
Let $x \in \bigcup_{i=1}^{\infty} \bigcap_{k=i}^{\infty} F_{\overline{k}}$, i.e. there exists $N \in \N$ such that $x \in  F_{\overline{k}}$ for every integer $k \geq N$, and suppose on the contrary that $x \notin \Lex_{G}$. This implies that there exists a non-trivial $g \in G$ such that $gx \lexx x$. By Lemma~\ref{lemma:Monotonicity-Lex} there exists $N' \in \N$, such that for every integer $k \geq N'$, we have that $\overline{k}^{t} gx > \overline{k}^{t} x$. But this is absurd because $x \in F_{\overline{k}}$ for $k \geq \max \{ N , N' \}$. The other inclusion is analogously derived from Lemma~\ref{lemma:Monotonicity-Lex}.

Furthermore, observe that $\left( \bigcap_{k=i}^{\infty} F_{\overline{k}} \right)_{i=1}^{\infty}$ is a non-decreasing nested sequence of convex sets, which implies that $\Lex_G$ is also convex. Finally, since the pair $(\R^{n}, \succeq)$ is a total order, then $\Lex_{G}$ must be a fundamental set for $G \leq S_{n}$. \qed 
\end{proof}

We now show that $\overline{\Lex_{G}}$, the (topological) closure of $\Lex_{G}$, is a fundamental domain for any $G \leq S_{n}$. %Moreover, it must be a polyhedral cone by~\ref{proposition:fd_generators}.

% This is actually the case, which implies that $\overline{\Lex_{G}}$ is a polyhedral cone by~\ref{proposition:fd_generators}.

\begin{theorem}\label{theorem:Lex-is-FD}
For any $G \leq S_{n}$, the closure of $\Lex_{G}$ is a fundamental domain.
\end{theorem}

\begin{proof}
Let $G \leq S_{n}$. We want to show that the non-empty closed set $\overline{\Lex}_{G} \subseteq \R^{n}$ satisfies:
\begin{enumerate}[label={(\roman*)}]
	\item $\bigcup_{g \in G} g \overline{\Lex}_{G} = \R^{n}$,
	\item $ g \inte( \overline{\Lex}_{G} )  \bigcap \inte ( \overline{\Lex}_{G}) = \emptyset, \,\, \forall g \in G \setminus \{ \id \}$.
\end{enumerate}
Notice that (i) follows directly as $\bigcup_{g \in G} g \Lex_{G} = \R^{n}$ since $\Lex_G$ is a fundamental set.
% First, we prove that $\overline{\Lex}_{G}$ satisfies (i). Note that by definition $\bigcup_{g \in G} g \overline{\Lex}_{G} \subseteq \R^{n}$. For the converse, let $x \in \R^{n}$. If $x \in \Lex_{G}$, then $x \in \Lex_{G} \subseteq \overline{\Lex}_{G}$ and we are done. If $x \notin \Lex_{G}$, then there exists a non-trivial $g \in G$ such that $gx \in \Lex_{G}$. Then $x \in g^{-1}\Lex_{G} \subseteq g^{-1} \overline{\Lex}_{G} $ because $g^{-1}$ is an orthogonal linear transformation.

For (ii), note that $\inte (\overline{\Lex}_{G}) = \inte (\Lex_{G})$ because $\Lex_{G}$ is a convex set by Lemma~\ref{lemma:ss_liminf_lex}. So it suffices to prove $ g\inte(\Lex_{G} ) \cap \inte ( \Lex_{G}) = \emptyset$ for every non-trivial $g \in G$. Suppose on the contrary that there exists a non-trivial $g \in G$ such that $x \in  g(\inte(\Lex_{G} )) \cap \inte ( \Lex_{G})$. Hence, $x$ and $g^{-1}x$ belong to $\inte ( \Lex_{G})\subseteq \Lex_{G}$. This is clearly absurd as $\preceq$ is a total order.\qed
%i.e. there exist $\epsilon_{1}, \epsilon_{2} > 0$ such that the balls $B_{\epsilon_{1}}(x), B_{\epsilon_{2}}(gx) \subseteq \Lex_{G}$ . This implies that for every $i \in [n]$, $x_{i} = x_{g^{-1}(i)}$. Now, let $0 < \epsilon < \epsilon_{1}$, ${j := \min(\supp(g))}$, where $\supp(g)$ the indices in $[n]$ moved by $g$. For $i \in [n]$ define
% \begin{equation*}
% \hat{x}_{i} = \left\{ \begin{array}{ll}
% x_{i} - \epsilon / 2 & \mbox{if     } i = j, \\
% x_{i} & \mbox{if    } i \neq j.
% \end{array} \right.
% \end{equation*} 
% This is absurd because $\hat{x} \in B_{\epsilon_{1}}(x)$ and $\hat{x} \notin \Lex_{G}$.\qed 
\end{proof}

% \begin{corollary}\label{section_ss_lexpolyhedron}
% Let $G \leq S_{n}$. Then the closure of $\Lex_{G}$ is a polyhedron.
% \end{corollary}
% \begin{proof}
% Recall that $\Lex_{G}$ is convex. Then by \ref{theorem:Lex-is-FD} we have that $\overline{\Lex}_{G}$ is a convex fundamental domain for $G$, hence it is a polyhedron by \ref{proposition:fd_generators}.\qed 
% \end{proof}

\subsubsection{A Characterization of \texorpdfstring{$\overline{\Lex}_{G}$ using the Schreier-Sims Table}.}\label{subsection_ss}

In what follows, we provide a characterization of $\overline{\Lex}_G$, which in particular allows to compute its facets efficiently. Indeed, we show that its description coincides with the Schreier-Sims inequalities for $G \leq S_{n}$~\cite{salvagnin18} (where computational results can be found). 

% Cut part ~Schreier-Sims

The \emph{Schreier-Sims table} is a representation of a permutation group $G \leq S_{n}$. The construction is as follows. Consider the chain of nested pointwise stabilizers defined as: $G^{0}:= G$ and $G^{i} := \{g \in G^{i-1} \talque g(i) = i \}$ for each $i\in[n]$.
Note that the chain is not necessarily strictly decreasing (properly), and we always have that ${G^{n-1} = \{ \id \}}$. For a given $i\in [n]$ and  $j\in\text{Orb}_{G^{i-1}}(i)$, let  $h_{i, j}$ be any permutation in $G^{i-1}$ which maps $i$ to $j$. Hence, $U_{i} := \{h_{i, j}: j \in\text{Orb}_{G^{i-1}}(i)\}$ is a transversal for the cosets of $G^i$ in $G^{i-1}$.

We arrange the permutations in the sets $U_{i}$, for $i \in [n]$, in an $n \times n$ table $T$ where $T_{i,j}= h_{i, j}$ if $j \in \Orb_{G^{i-1}}(i)$ and $T_{i,j}=\emptyset$ otherwise.

The most interesting property of this construction is that each $g \in G$ can be uniquely written as $g = g_{1} g_{2} \cdots g_{n}$ with $g_{i} \in U_{i}$, for $ i \in [n]$. Therefore, the permutations in the table form a set of generators of $G$ which is called a \emph{strong generating set (SGS)} for $G$~\cite{seress03}. %Equation \ref{sgs_equation} shows that $g \in G$ can be expressed as a product of at most $n$ permutations in the set. Indeed, the permutations $g_{2}, \dots, g_{n}$ all stabilize element $1$, forcing $g_{1}$ to be $T_{1, g(1)}$. Then, as $g_{3}, \dots, g_{n}$ all stabilize element $2$, we must have $g_{1} g_{2}(2) = g(2) \iff g_{2}(2) = g_{1}^{-1}g(2)$, and thus $g_{2} = T_{2, g_{1}^{-1}g(2)}$. A similar reasoning yields $g_{3}, \dots, g_{n}$. Given a set of generators $S$ for $G$, this powerful set of generators can be computed in time $O(n^{2} \log^{3} |G| + |S| n^{2} \log |G| )$ \cite[Theorem 4.2.4]{seress03}.

The \emph{Schreier-Sims polyhedron}, denoted by $\SSP_{G}$, is the polyhedron given by the inequalities $x_{i} \geq x_{j}$ for all $T_{i,j} \neq \emptyset$. Theorem~\ref{theorem:sp_lex} states that $\overline{\Lex}_{G} = SS_{G}$. A crucial observation to prove this is that for any vector $x \in \R^{n}$, $x$ is in the closure of $\Lex_{G}$ if and only if $x$ can be perturbed into the interior of $\overline{\Lex}_{G}$, where the perturbed vector is lexicographically maximal in its orbit.

% \begin{remark}
% Note that a bound for the number of inequalities of $\SSP_{G}$ is $\frac{n^{2} - n}{2}$. Shortly we will prove a tight linear upper bound in \ref{subsection_sstree_facets}.
% \end{remark}

The following lemma characterizes the boundary of $\overline{\Lex}_{G}$ in terms of a special perturbation which we call ``tie-breaker" perturbation, because it breaks all possible ties between the vector's entries. We will use this lemma to prove that the facets of $\overline{\Lex}_{G}$ are in fact Schreier-Sims inequalities.

\begin{lemma}\label{lemma:tiebreaker}
Let $G \leq S_{n}$, $x \in \R^{n}$, and $0 < \epsilon < M$, where
\begin{equation}\label{M}
M:= \left\{ \begin{array}{ll}
1 & \text{if $x_{i} = x_{j}, \squad \forall i, j \in [n]$}, \\
\min \{ | x_{i} - x_{j} | > 0 \talque i,j \in [n] \} & \text{otherwise}.
\end{array} \right.
\end{equation}
We define the $\epsilon$-tie-breaker perturbation for $x$ as
\begin{equation*}
x_{i}^{\epsilon} := x_{i} - \frac{i \epsilon}{n^{2}} \quad \text{for $ i \in [n]$.}
\end{equation*}
Then $x$ is in $\overline{\Lex}_{G}$ if and only if for any $0 < \epsilon < M$, $x^{\epsilon}$ belongs to $\Lex_{G}$.
\end{lemma}
\begin{proof}
Let $x \in \R^{n}$ and $M$ be defined as above and suppose that $x^{\epsilon} \in \Lex_{G}$ for all $0 < \epsilon< M$. Since $\lim\limits_{\varepsilon \to 0} x^{\epsilon} =x$ then $x\in \overline{\Lex}_{G}$.

For the converse, suppose $x \in \R^{n}$ and let $0 < \epsilon < M$, with $M$ defined as in~\eqref{M}. %and consider the $\epsilon$-tie-breaker perturbation for $x$
% \begin{equation*}
% x_{i}^{\epsilon} := x_{i} - \frac{i \epsilon}{n^{2}} \quad \text{for $ i \in [n]$,}
% \end{equation*}
Notice that in $x^{\varepsilon}$ ties in $x$ are broken without changing the relative order of its coordinates, that is, if $x_i<x_j$ then $x_i^{\epsilon} < x_j^{\epsilon}$. Also, $x_i^{\epsilon}\neq x_j^{\epsilon}$ for $i\neq j$.
Suppose that $x^{\epsilon}$ is not in $\Lex_{G}$. We want to show that $x$ does not belong to $\overline{\Lex}_{G}$. If $x^{\epsilon} \notin \Lex_{G}$, there exists a non-trivial $g \in G$ such that $gx^{\epsilon} \lexx x^{\epsilon}$. Let us characterize this $g$. %Note that because all coordinates of $x^\epsilon$ are different, the set of fixed indices $[n] \setminus \supp(g)$ coincides with the set of indices whose coordinates are stabilized by $g$, i.e. $\fix_{x^{\epsilon}}(g) := \{ i \in [n] \talque x^{\epsilon}_{g^{-1}(i)} = x^{\epsilon}_{i} \}$. 
Let $i \in [n]$ denote the (largest) length of the prefix that~$g$ fixes in $x^{\epsilon}$, i.e., $(gx^{\epsilon})_{j} = x^{\epsilon}_{j}$ for $j \leq i$ and $(gx^{\epsilon})_{i+1} \neq x^{\epsilon}_{i+1}$. Hence, as the coordinates of $x^{\epsilon}$ are pairwise different, $g$ belongs to the subgroup of $G$ that fixes indices $1, \dots, i$ pointwise, i.e. $g \in G_{([i])}$. Then because $g$ improves $x^{\epsilon}$ lexicographically and $g$ fixes every $j \leq i$, the improvement should occur from entry $i + 1$ onwards.  
Hence, as $gx^{\epsilon} \lexx x^{\epsilon}$, it must hold that $(gx^{\epsilon})_{i+1} = x^{\epsilon}_{g^{-1}(i+1)} > x^{\epsilon}_{i+1}$ and $g^{-1}(i+1)>i + 1$.

%Recall that all coordinates are different so $g^{-1}(i+1) > i + 1$ implies $x^{\epsilon}_{g^{-1}(i+1)} > x^{\epsilon}_{i+1}$, i.e., the improvement occurs exactly at entry $i+1$. 
By construction of our $\epsilon$ perturbation, it must also hold that $x_{g^{-1}(i+1)} > x_{i+1}$. Indeed, it cannot hold that $x_{g^{-1}(i+1)} = x_{i+1}$, since the perturbation is increasing in the vector's indices and $g^{-1}(i+1) > i + 1$. Neither it can happen that $x_{g^{-1}(i+1)} < x_{i+1}$ since this implies that $x_{g^{-1}(i+1)}^{\epsilon} < x_{i+1}^{\epsilon}$. In consequence,
it holds that $gx \succ x$ and hence $x\not\in \Lex_G$. Moreover, the ball $B_{M/2}(x) \subseteq \Lex_G^{c}$, as $g$ fixes the indices $1, \dots, i$ pointwise and exchanges $x_{i+1}$ for a strictly greater entry $x_{g^{-1}(i+1)}$ in $x$, where $g^{-1}(i+1)>i+1$.\qed%, cannot be outruled by any perturbation $\overline{\epsilon} \leq \epsilon$.\qed
\end{proof}

Now we are ready to show a characterization of $\overline{\Lex}_{G}$ by an explicit set of inequalities.

\begin{theorem}\label{theorem:sp_lex}
Let $G \leq S_{n}$. Then $\overline{\Lex}_{G} = \SSP_G$.
\end{theorem}
\begin{proof}
% Suppose on the contrary that $x \notin \SSP_{G}$ but $x \in  \overline{\Lex}_{G}$, i.e. $x' \in \Lex_G$ by \ref{lemma:tiebreaker}. Then there exists a minimal $i \in [n]$ and $j \in \Orb_{G_{i-1}}(i)$, such that $x_{i} < x_{j}$ where $i < j$. Then, exists $g \in G_{i-1}$ such that $gx \succ x$ which is a contradiction.
Suppose on the contrary that $x \notin \SSP_{G}$ but $x \in  \overline{\Lex}_{G}$, i.e., $x^{\epsilon} \in \Lex_G$ for all $0<\epsilon<M$ by Lemma~\ref{lemma:tiebreaker}. As $x \notin \SSP_{G}$, there exists a minimal $i \in [n]$ and $j \in \Orb_{G_{i-1}}(i)$, such that $x_{i} < x_{j}$ where $i < j$, and thus $x_{i}^{\epsilon} < x_{j}^{\epsilon}$. Then, there exists $g \in G_{i-1}$ such that $gx^{\epsilon} \succ x^{\epsilon}$ which is a contradiction.

For the converse, suppose $x \in \SSP_{G}$. If for each index-orbit $\Orb\subseteq [n]$, all the components of $x$ indexed by $\Orb$ are different, then $x \in \Lex_{G}$ since for any pair $(i,j)$ such that $i \in [n]$, $j \in \Orb_{G_{i-1}}(i)$, and $i < j$, the corresponding Schreier-Sims inequality is strict, i.e. $x_{i} > x_{j}$. If not, for every coordinate-tie within an orbit apply the tie-breaker perturbation. Therefore, the perturbed vector belongs to $\Lex_G$, i.e., $x \in \overline{\Lex_{G}}$ by Lemma~\ref{lemma:tiebreaker}. \qed
\end{proof}

The next result exhibits the generality of our GDD method for constructing fundamental domains. It shows that by choosing $\alpha_i$ as the canonical basis vectors in our GDD construction the algorithm outputs $\SSP_{G}$. We note that this also gives an alternative proof to Theorem~\ref{theorem:Lex-is-FD}.

\begin{proposition}
For any group $G \leq S_{n}$ the set $\SSP_{G}$ is a GDD.% cref{algo:GeneralizedDirichletDomain} can construct the Schreier-Sims fundamental domain in polynomial-time.
\end{proposition}
\begin{proof}
We show that $\SSP_{G}$ can be obtained from Algorithm~\ref{algo:GeneralizedDirichletDomain} by choosing $\alpha_i$ equal to the canonical vectors. We begin the procedure with $\alpha_{1} := e_{1}$, then $G_{1}$, in Line~\ref{algo_compute} of Algorithm~\ref{algo:GeneralizedDirichletDomain}, corresponds to the pointwise stabilizer of the index $1 \in [n]$. Hence, if in iteration $i \in [n]$ we choose $\alpha_{i} := e_{i}$, then the subgroup $G_{i}$ which stabilizes $\alpha_{1}, \dots, \alpha_{i}$ is equal to $G_{([i-1])}$. After at most $n$ iterations we have that $F = \SSP_{G}$. \qed %The polynomial-time complexity of this construction follows after noting that computations in lines \ref{algo_compute} and \ref{algo_choose} correspond to the rows of the Schreier-Sims table which can be computed in polynomial-time \cite[Thm. 4.2.4.]{seress03}.\qed 
\end{proof}

% \subsubsection{The Schreier–Sims Tree}\label{subsection_tree}

As the Schreier-Sims table has $O(n^2)$ many entries, the number of facets of the Schreier-Sims fundamental domain is at most $O(n^2)$. In what follows we show a tighter bound of $O(n)$. To this end, we notice that several of the added inequalities are redundant. %We introduce \emph{Schreier-Sims trees}, which characterize the facets of the Schreier-Sims polyhedron $\SSP_G$. More precisely, in \ref{ss_forest}, we construct a forest for $G$, where every $G$-orbit in $[n]$ is associated to a Schreier-Sims tree. This construction is helpful to determine the number of facets of $\SSP_G$. %Nevertheless, it sheds light on its seemingly weak symmetry breaking power, as two non-isomorphic groups can have the same tree representation. 
\begin{theorem}\label{subsection_sstree_facets}
Let us consider a group $G \leq S_{n}$ and let $f$ denote the number of $G$-orbits in $[n]$. Then $\SSP_{G}$ is a polyhedron with at most $n-f$ facets.
\end{theorem}
\begin{proof}
Let $D=([n],E)$ be a directed graph defined as follows. For each $i\in [n]$ we have that $(i,j)\in E$ for each $j\in \Orb_{G^{i-1}}(i)$. By construction, $D$ is a topological sort, and hence it is a directed acyclic graph (DAG). 

\medskip
\noindent{\textit{Claim}:} Let $j\in [n]$. If $(i,j),(k,j)\in E$ then either $(i,k)\in E$ or $(k,i)\in E$.
\medskip

Indeed, without loss of generality, let us assume that $i<k$. As $(i,j)\in E$ then $j\in \Orb_{G^{i-1}}(i)$. Similarly, it holds that $j\in \Orb_{G^{k-1}}(k)\subseteq \Orb_{G^{i-1}}(k)$. Therefore, by transivity, $k\in \Orb_{G^{i-1}}(i)$, and hence $(i,k)\in E$. This shows the claim.

Let $\tilde{D}=([n],\tilde{E})$ be the \emph{minimum equivalent graph} of $D$, that is, a subgraph with a minimum number of edges that preserves the reachability of $D$. Hence, there exists a $(u,v)$-dipath in $D$ if and only if there exist a $(u,v)$-dipath in $\tilde{D}$. Notice that 
\[\SSP_{G}=\{x\talque x_i\ge x_j \text{ for all }  (i,j)\in E\}.\]
We define \[\widetilde{\SSP}_{G} = \{x\talque x_i\ge x_j \text{ for all }  (i,j)\in \tilde{E}\},\]
then $\SSP_{G}=\widetilde{\SSP}_{G}$. 
Clearly we have that $\SSP_{G}\subseteq \widetilde{\SSP}_{G}$. On the other hand, if $x_i\ge x_j$ is an inequality of $\SSP_{G}$, then there exists an $(i,j)$-dipath in $\widetilde{\SSP}_{G}$ and hence $x_i\ge x_{i_1}\ge x_{i_2} \ge \ldots x_{i_k} \ge x_j$ is a valid set of inequalities for $\widetilde{\SSP}_{G}$, for certain nodes $i_1,\ldots,i_k$. We conclude that $\SSP_{G}=\widetilde{\SSP}_{G}$.

No we argue that $\tilde{D}$ is a collection of at least $f$ out-trees. Indeed, lets assume by contradiction that for $j\in [n]$ there exists two distinct nodes $i,k$ such that $(i,j),(k,j)\in \tilde{E}$. By our previous claim, $k$ is reachable from $i$ in $D$ (or analogously $i$ is reachable from $k$), and hence the same is true in $\tilde{D}$. This is a contradiction as the edge $(i,j)$ could be removed from $\tilde{D}$ preserving the reachability. As $\tilde{D}$ is a DAG, then $\tilde{D}$ must be a collection of node-disjoint out-trees. Finally, note that the smallest element in each orbit of $G$ in $[n]$ has in-degree $0$ in $D$, and hence also in $\tilde{D}$. Therefore $\tilde{D}$ has at least $f$ different trees, which implies that $\tilde{D}$ has at most~$n-f$ edges.
\qed
\end{proof}
% \begin{proof}
% If $G$ is transitive, then $f=1$ and $W_{G}$ is a Schreier-Sims tree with $n$ nodes. Then $W_{G}$ has $n-1$ edges. If $G$ is not transitive, then $W_{G}$ is a forest with $f$ connected components. Then it has $n-f$ edges, i.e. $\overline{\Lex}_{G}$ has $n-f$ facets. \qed 
% \end{proof}

This means that every permutation group admits a %n exact 
fundamental domain with at most $n-1$ facets. We complement this theorem by the following observation.

% \begin{proposition}
% There exists a $G \leq S_{n}$ for which every exact fundamental domain has $\Omega(n)$ facets.
% \end{proposition}

\begin{proposition}
Any fundamental domain for $S_{n}$ has $n-1$ facets.
\end{proposition}

\begin{proof}
Since $S_{n}$ is generated by the transpositions $(i \,\, j)$ for all $i \neq j \in [n]$ which correspond to reflections with reflection axis $x_{i} = x_{j}$, we conclude that $S_{n}$ is a reflection group. Moreover, since fundamental domains for reflection groups are unique (up to actions of the group), see Coxeter~\cite[pp. 79 -- 81]{coxeter48}, any fundamental domain for $S_{n}$ is equivalent to the Schreier-Sims fundamental domain for $S_{n}$. This symmetry breaking set has $n-1$ facets and can be described by the inequalities $x_{i} \geq x_{i+1}$ for every $i \in [n-1]$. \qed
\end{proof}

\section{Overrepresentation of Orbit Representatives}\label{section_mul-uni}

A desirable property of symmetry breaking polyhedra is that they select a unique representative per $G$-orbit. In general, the definition of fundamental domains only guarantees this for vectors in their interior. Recall that a subset $R$ of $\R^{n}$ which contains exactly one point from each $G$-orbit is called a \emph{fundamental set}. The following result shows that closed convex fundamental sets are only attained by reflection groups. In other words, the only groups that admit fundamental domain containing unique representatives for every orbit are reflection groups.

\begin{theorem}\label{theorem:reflexion-unique}
Let $G \leq O_{n}(\R)$ finite. Then $G$ admits a fundamental domain $F$ with $|F\cap O|=1$ for every $G$-orbit $O\subseteq \mathbb{R}^n$ if and only if $G$ is a reflection group.
\end{theorem}

\begin{proof}
Suppose $G$ admits a closed convex fundamental set $F \subseteq \R^{n}$, i.e. $F$ is a fundamental domain and for every $x \in \R^{n}$ we have that
\begin{equation*}
    \Orb_{G}(x) \cap F = \{ gx \},
\end{equation*}
for some $g \in G$. By Proposition~\ref{proposition:fd_generators} we know that $F$ is a polyhedral cone and we can write it as
\begin{equation*}
    F = \bigcap_{g\in A} H_{g},
\end{equation*}
where $A$ is a generating set for $G$. We want to show that $A$ is a set of reflections.

Let $g$ be a non-trivial element of $A$ and consider its associated half-space~$H_{g}$. We know that $H_{g}^{=}$ is a supporting hyperplane for $F$, and $F \cap gF$ has dimension $n-1$. Suppose $x$ is an arbitrary vector in $F \cap gF$. Then $g^{-1}x \in F$, and hence $gx = x$ because $F$ contains a unique representative of $x$. Now consider the span of $F \cap gF$ and let $\hat{g}$ denote the restriction of the orthogonal transformation $g$ to this linear subspace. As $\hat{g}$ fixes every point in the relative interior of $F \cap gF$, which is $n-1$ dimensional, we have that $\hat{g}$ acts trivially in $\text{span} (F \cap gF)=H_{g}^{=}$. Since $g$ is a non-trivial isometry, every vector $y\in (H_{g}^{=})^{\perp}$ must satisfy that $gy=-y$. We conclude that $g$ is a reflection with respect to the hyperplane $H_{g}^{=}$. For the converse implication, see Coxeter~\cite[pp. 79 -- 81]{coxeter48} and notice that his construction gives an exact fundamental domain as its facets are defined by reflections.\qed
\end{proof}

As a corollary we can characterize when the fundamental set $\text{Lex}_G$ is closed. % (\ref{section_ss_G_closed}).
Alternatively, this characterizes when $\SSP_{G}$ contains a unique representative for \emph{every} orbit. Equivalently, this characterizes the groups for which the fundamental set $\Lex_{G}$ is closed. Our proof utilizes the next lemma which provides an orthogonal decomposition of $\Lex_{G}$ when $G$ is a direct product.

% This is actually the case, which implies that $\overline{\Lex_{G}}$ is a polyhedral cone by~\ref{proposition:fd_generators}.

\begin{lemma}\label{Lemma:Lexprod}
Consider $G \leq S_{n}$. Assume that $O_{1}, O_{2}$ is a partition of $[n]$, and $G_{i} \leq S_{O_{i}}$ for $i \in \{1,2\}$. If ${G = G_{1} \times G_{2}}$, then $\Lex_{G} = \Lex_{G_{1}} \times \Lex_{G_{2}}$.
\end{lemma}
\begin{proof}
For $x\in\R^n$ and $S\subseteq [n]$, let us denote by $x_{S}$ the vector $x$ restricted to the coordinates in $S$. Let us also denote $x_{O_i} \succ_i y_{O_i}$ if $x_{O_i}$ is lexicographically larger than $y_{O_i}$ (without altering the order of elements in~$O_i$). Also, for $(g_1,g_2)\in G_1\times G_2$, we denote by $x\rightarrow (g_1,g_2)x$ the action where $g_1$ permutes the coordinates in $O_1$ and $g_2$ permutes the coordinates in $O_2$.

We must show that the following are equivalent:
\begin{itemize}
    \item[(i)] $x \succeq gx$ for all $g\in G_1\times G_2$.
    \item[(ii)] $x_{O_i}\succeq_{i} g_i x_{O_i}$ for $i\in\{1,2\}$ for every $(g_1,g_2)\in G_1\times G_2$.
\end{itemize}
Clearly, (ii) is equivalent to $x\succeq (g_1,id) x$ and $x\succeq (id,g_2) x$ for every $(g_1,g_2)\in G_1\times G_2$. This last condition is necessary for (i). To see that is also sufficient, assume that $x\prec (g_1,g_2) x$ for some $(g_1,g_2)\in G_1\times G_2$. Let $i$ be the first coordinate where $x_i < ((g_1,g_2) x)_i$. Let us assume that $i\in O_1$ (the case $i\in O_2$ is analogous), and hence $x_j= (g_1 x_{O_1})_j$ for all $j<i$ such that $j\in O_1$. This implies that $x_{O_1}\prec_1 g_1 x_{O_1}$. The lemma follows.
\qed
\end{proof}
% \begin{proof}
% Without loss of generality suppose that $O_{1}, O_{2}$ is an ordered partition of $[n]$, i.e. that the first $| O_{1} |$ integers are in $O_{1}$ and the rest are in $O_{2}$, because if it is not ordered a simple relabeling suffices. Let $\pi_{i}: \R^{|O_{1}|} \times \R^{|O_{2}|} \to \R^{|O_{i}|}$ be the canonical projection of $x\in \R^{n}$ onto $\R^{|O_{i}|}$. Then for  every $x \in \R^{n}$ we denote $x_{i} := \pi_{i}(x)$, i.e. $x = (\pi_{1}(x), \pi_{2}(x))$, and the product-action of $G_{1} \times G_{2}$ on $\R^{n}$ is naturally given by $g(x) := (g_{1}(x_{1}), g_{2}(x_{2}))$, such that $g = (g_{1}, g_{2})$, with $g_{i} \in G_{i}$.

% Now, let $x \in \Lex_{G}$. Suppose on the contrary that $x_{2} \notin \Lex_{G_{2}}$. Given that $G$ contains every permutation of the form $(\id_{1}, g_{2})$ for every $g_{2} \in G_{2}$, if $g^{*}_{2}(x) \in \Lex_{G_{2}}$, then $g(x) \succ x$ with $g := (\id_{1}, g_{2}^{*})$, which is absurd. This implies that $\Lex_{G} \subseteq \Lex_{G_{1}} \times \Lex_{G_{2}}$. For the converse, let $x \in \Lex_{G_{1}} \times \Lex_{G_{2}}$ and suppose $x$ is not in $\Lex_{G}$, i.e. there exists $g = (g_{1}, g_{2})$ in $G$ such that $g(x) \succ x$. As $x_{i} \in \Lex_{G_{i}}$, then $x_{i} \succeq g_{i}(x_{i})$. Then $x = (x_{1}, x_{2}) \succeq (g_{1}(x_{1}), g_{2}(x_{2}))$ which is a contradiction. \qed 
% \end{proof}

\begin{corollary}\label{section_ss_G_closed}
Let $G \leq S_{n}$. Suppose that $G$ partitions $[n]$ into a collection of orbits $O_{1}, \dots, O_{m} \subseteq [n]$, $m \in [n]$. Then $\Lex_{G}$ is closed if and only if
\begin{equation*}
G = S_{| O_{1} |} \times \cdots \times S_{| O_{m} |}.
\end{equation*}
\end{corollary}
\begin{proof}%[\ref{section_ss_G_closed}]
Suppose that $G \leq S_{n}$ is transitive. Given that the only transitive reflection group of $S_{n}$ is $S_{n}$ itself, Theorem~\ref{theorem:reflexion-unique} implies that if $G$ is a proper transitive group of $S_{n}$ then $G$ does not admit a closed convex fundamental set. Hence $\Lex_{G}$ cannot be closed since $\Lex_{G}$ is a convex fundamental set for any $G \leq S_{n}$.

Suppose $G$ is not transitive on $[n]$. Then the if part is straightforward by Lemma~\ref{Lemma:Lexprod} and Theorem~\ref{theorem:reflexion-unique}. For the converse, without loss of generality assume that $G$ is a subgroup of $S_{| O_{1} |} \times \cdots \times S_{| O_{m} |}$ and that the orbits are reordered (relabeled) as $\{1, \dots, n_{1} \}$, $\{n_{1} + 1, \dots, n_{2} \}$, \dots, $\{n_{m-1} + 1, \dots, n\}$, where $|O_{1}|= n_{1}$, ${|O_{2}| = n_{2} - n_{1}}$, \dots, ${ |O_{m-1}| = n_{m-1} - n_{m-2} }$, $|O_{m} | = n - n_{m-1}$. We perform this reordering without changing the order of the variables in the lexicographic order, so that $\Lex_G$ is maintained.

Now, suppose that $\Lex_{G}$ is closed, i.e. $G$ is a group generated by reflections, and that $G \leq S_{| O_{1} |} \times \cdots \times S_{| O_{m} |}$. Note that if $g \in G$ is a reflection, then it must be a transposition $(i \, \,j)$ for some $i,j \in [n]$. Indeed, consider the decomposition of $g$ into disjoint cycles $\{c_{1}, \dots, c_{L}\}$, i.e. $g = c_{1} \cdots c_{L}$, and note that the invariant subspace of any cycle $c_{l}$ is given by the equalities $x_{i} = x_{j}$ for all $i,j$ moved by $c_{l}$. Since the invariant subspace of $g$ is $n-1$, then $g$ must be equal to a single transposition $(i \,\, j)$. Therefore, $G$ is generated by transpositions, and since it acts transitively on each $G$-orbit $O_{k}$, then the action of $G$ restricted to $O_{k}$ is isomorphic to $ S_{| O_{k} |}$. Moreover, the direct product follows after noting that every $g$ is a product of cycles of order $2$, hence $S_{| O_{k} |} \subseteq G$ for every $k \in [m]$.
\qed
\end{proof}

% \begin{proof}
% Suppose that $G \leq S_{n}$ is transitive. Given that the only transitive reflection group of $S_{n}$ is $S_{n}$ itself, then \ref{theorem:reflexion-unique} implies that if $G$ is a proper transitive group of $S_{n}$ then $G$ does not admit a closed convex fundamental set. Hence $\Lex_{G}$ cannot be closed since $\Lex_{G}$ is a convex fundamental set for any $G \leq S_{n}$. \qed
% \end{proof}

In integer programming problems, we are concerned about the number of representatives of binary orbits in a fundamental domain. The Schreier-Sims domain can be weak in this regard, as shown in the following example. 

\begin{example}\label{ex:manyreps}
Let $n \in \N$ be divisible by $3$, and consider the direct product 
\[
    G := C^{1} \times C^{2} \times \cdots \times C^{n/3},
\]
where $C^{i}$ for $i \in [n/3]$ is the cyclic group on the triplet 
\[
    (3(i-1) + 1, \, 3(i-1) + 2, \, 3(i-1) + 3).
\]
Consider the binary vector $x := (1,1,0,1,1,0, \dots, 1,1,0)$. For each vector in the $G$-orbit of $x$, there are three possible values for each triplet: 
\begin{equation*}
1,1,0 \squad \text{ or } \squad 0,1,1 \squad \text{ or } \squad 1,0,1. 
\end{equation*}
Therefore, the orbit of $x$ has cardinality $3^{n/3}$. The fundamental domain $\SSP_{G}$ for $G$ can be described as follows 
\begin{equation*}
\SSP_{G} = \{ x \in \R^{n} \talque x_{3(i-1) + 1} \geq x_{3(i-1) + 2} \text{ and } x_{3(i-1) + 1} \geq x_{3(i-1) + 3},\,\,\forall i  \in [n/3]  \}.
\end{equation*}
It is clear that each vector in $\SSP_{G} \cap \Orb_{G}(x)$ admits two options for its index triplets: $1,1,0$ and $1,0,1$. As a result $|\SSP_{G} \cap \Orb_{G}(x)| = 2^{n/3}$. \xqed
\end{example}

We propose a definition for a theoretical classification of symmetry breaking systems inspired by our findings, and we consider two attributes to rank symmetry breaking systems: the complexity of their separation, and their symmetry breaking power, i.e. their effectiveness to cut isomorphic points. These two properties seem to be a longstanding trade-off in mathematical programming with respect to symmetry breaking systems \cite{margot10,hojny+pfetsch18}, and this trade-off has also been recognized in \emph{constraint satisfaction problems}~\cite{handbookcp06}. In the latter, it is challenging to identify a symmetry breaking system which is both \emph{effective}, in the sense that it rules out a large portion of the search space, and \emph{compact}, which means that the symmetry breaking inequalities can be checked in a reasonable amount of time~\cite{shlyakhter07,crawford+etal96}.

Let $X$ be a $G$-invariant subset of $\R^{n}$ (e.g. $X=\{0,1\}^n$). Let $\mathcal{O}(G,X)$ be the set of all $G$-orbits in $X$. Motivated by our previous discussion, we define, for a fixed $G$, the \emph{worst-case effectiveness of $F$ on $X$} as
\begin{equation*}
    \Lambda_{G,X}(F) := \max_{O \in \mathcal{O}(G,X)} | F \cap O |.
\end{equation*} 
Now, we use our GDD algorithm to obtain a suitable fundamental domain in Example~\ref{ex:manyreps} with $\Lambda_{G,\{0,1\}^n}(F)=1$ while $\Lambda_{G,\{0,1\}^n}(\SSP_G)=2^{\Omega(n)}$. %Recall that the Schreier-Sims fundamental domain can have a very weak performance as a symmetry breaker with $\Lambda_{G, \{0,1\}^{n}}(\SSP_{G}) = \Omega(2^{\Omega(n)})$. We now construct, for the same group $G$, a fundamental domain with polynomially many facets $F$ such that $\Lambda_{G, \{0,1\}^{n}}(F) = 1$.

\newenvironment{contexample}{
   \addtocounter{example}{-1} \begin{example}[continued]}{
   \end{example}}
\begin{contexample}   
We construct a GDD $F$ with $\Lambda_{G, \{0,1\}^{n}}(F) = 1$. First, note that $G$ has $n/3$ orbits in $[n]$ given by:
\[
    \Delta_{i} := \{ 3(i-1) + 1, 3(i-1) + 2, 3(i-1) + 3 \}
\]
for $i \in [n/3]$. Therefore the following vectors, and their associated stabilizers, construct a generalized Dirichlet domain
\begin{align*}
\alpha_{1} &= (4,2,1, 0,0,0, 0, \dots, 0), \quad G_{\alpha_{1}}  = (C_{3})^{n/3 - 1} \\
\alpha_{2} &= (0,0,0, 4,2,1, 0, \dots, 0), \quad G_{\alpha_{2}}  = (C_{3})^{n/3 - 2}\\
\vdots \\
\alpha_{n/3} &= (0,0,0, 0, \dots, 0, 4, 2, 1), \quad G_{\alpha_{n/3}}  = \langle \id \rangle.
\end{align*}
such that $\Orb_{G}(x) \cap F = \{ x \}$ for any $x\in\{0,1\}^n$. The number of cosets in each iteration is 3. Omitting the trivial coset, the number of inequalities that defines our GDD is $2\cdot (n/3)$. \xqed
%In our GDD construction we choose for every $i \in [n/3]$ a vector $\alpha_i = (0^{3(i-1)},4,2,1,0^{n-3i})$, where $0^r$ is an $r$-dimension 0 vector. We obtain that  $\Orb_{G}(x) \cap F = \{ x \}$ . 
\end{contexample}

% \bigskip
% Conclusions: 1/2 page.
% \noindent\textbf{Future work.} 
\section{Future Work}
Our work leaves several major questions.

\medskip
\noindent\textbf{Q1:} Does our GDD construction exhaust all possible fundamental domains for a group of isometries, or are there other fundamental domains that are not GDDs? 
\medskip

Any light on this question can help creating new fundamental domains with potential practical relevance, or help us show impossibility results. This can also have consequences regarding our long term goal: understanding the tension (potentially trade-off) between the symmetry breaking effectiveness of a polyhedron and its complexity. A closely related question is whether we need the hypothesis of being subgroup consistent in Theorem~\ref{thm:FD_facet_subgroup_consistent}. If the answer to Q1 is positive, we would immediately conclude that Theorem~\ref{thm:FD_facet_subgroup_consistent} holds without assuming that the fundamental domain is subgroup consistent. %This also would help us determined wh%In this regards, a major line of work is given by the following question.

\medskip
\noindent\textbf{Q2:} Does every group of isometries admit a fundamental domain with a single representative of each binary orbit, and with a polynomial number of facets?
\medskip

It is not hard to imagine other interesting variants of this question. For example, we could be interested either in the extension complexity or complexity of the separation problem, instead of the number of facets. At the moment, the only information we have is that blindly choosing lexicographically maximal binary vectors as representatives should not help, as finding them is NP-hard~\cite{babai+luks83}. It is worth noticing that an answer to Q1 might help answering Q2, either positively or negatively. Alternatively, the relation between $\Lambda_{G,X}(F)$ and the number of facets of a fundamental domain $F$ is of interest, for example for $X=\{0,1\}^n$. On the other hand, we know that only reflection groups admit fundamental domains with $\Lambda_{G,\mathbb{R}^n}(F)=1$. Characterizing, for example, the class of groups that allows for $\Lambda_{G,\mathbb{R}^n}(F)=O(1)$ might also give us a better understanding on the limitations of symmetry breaking polyhedra.

% \begin{acknowledgements}
% We are greatly indebted to A. Behn, C. Hojny, M. Pfetsch, and V. Verdugo for fruitful discussions on the topic of this paper.
% \end{acknowledgements}

\subsubsection*{Acknowledgements} 

This work was partially funded by ANID/CONICYT Fondecyt Regular Nr. 1181527 and ANID – Millennium Science Initiative Program – NCN17\_059. Part of this work was done while the first author was affiliated to the University of O'Higgins, Chile. The third author was supported by Basal Program CMM-AFB 170001 from ANID (Chile). We are greatly indebted to A. Behn, C. Hojny, M. Pfetsch, and V. Verdugo for fruitful discussions on the topic of this paper. 

\bibliographystyle{spmpsci}
\bibliography{fd}
%\bibliography{symbre}

\end{document}